\PassOptionsToPackage{table}{xcolor}
\documentclass[10pt,twocolumn,letterpaper]{article}

\usepackage[table]{xcolor}
\usepackage[pagenumbers]{cvpr} 
\definecolor{cvprblue}{rgb}{0.21,0.49,0.74}

\usepackage{booktabs} 
\usepackage[pagebackref,breaklinks,colorlinks,allcolors=cvprblue]{hyperref}
\usepackage[ruled]{algorithm2e} 

\SetAlFnt{\small}
\SetAlCapFnt{\small}
\SetAlCapNameFnt{\small}
\SetAlCapHSkip{0pt}

\definecolor{firstplace}{rgb}{0.3,0.7,0.3}
\definecolor{secondplace}{rgb}{0.13,0.59,0.95}
\definecolor{thirdplace}{rgb}{1.0,0.76,0.03}
\definecolor{title_purple}{rgb}{0.65,0.1,0.65}
\definecolor{title_blue}{rgb}{0.05,0.05,0.95}

\newcommand{\mypara}[1]{\noindent \textbf{\textit{#1}.}}

\title{\textcolor{title_blue}{ACT-R}: \textcolor{title_blue}{A}daptive \textcolor{title_blue}{C}amera \textcolor{title_blue}{T}rajectories
 for Single-View 3D \textcolor{title_blue}{R}econstruction}

\newcommand{\equalcontrib}{\textsuperscript{*}}
\author{Yizhi Wang\equalcontrib$^{1}$ \and
Mingrui Zhao\equalcontrib$^{1}$ \\
{\tt \small$^1$Simon Fraser University}  \and
Hao Zhang$^1$ 
}

\begin{document}

\twocolumn[{
\maketitle
\vspace{-2.5em}
\centering
\includegraphics[width=.98\textwidth]{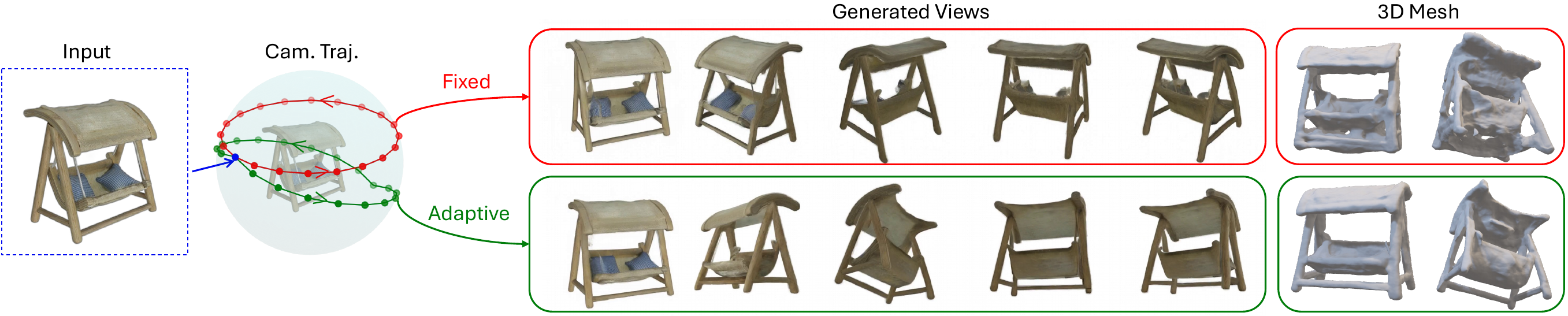}
\captionof{figure}{
ACT-R, for single-view 3D reconstruction, predicts an {\em adaptive camera trajectory} (green) to maximize the visibility of occluded object parts over a fixed sequence length (20 views). The trajectory, obtained in under 10s, is then used by a video generator (e.g., SV3D~\cite{voleti2024sv3d}) to produce a {\em sequence\/} of novel views for multi-view 3D reconstruction, here using NeUS~\cite{wang2021neus}.
Compared to a generic trajectory (red), at fixed elevation, ACT-R yields much cleaner results with more faithful recovery of occluded regions.
}
\vspace{1 em}
\label{fig:teaser}
}]
\renewcommand{\thefootnote}{}
\footnotetext[1]{$^*$Equal contribution.}

\begin{abstract}
We introduce the simple idea of {\em adaptive view planning\/} to multi-view synthesis, aiming to improve both occlusion revelation 
and 3D consistency for single-view 3D reconstruction. Instead of producing an unordered set of views independently or simultaneously, we 
generate a {\em sequence\/} of views, leveraging temporal consistency to enhance 3D coherence. Importantly, our view sequence is not determined by a pre-determined and fixed camera setup. Instead, we compute an {\em adaptive camera trajectory\/} (ACT), to maximize the visibility of occluded regions of the 3D object to be reconstructed. 
Once the best orbit is found, we feed it to a video diffusion model to generate novel views around the orbit, which can then be passed to any multi-view 3D reconstruction model to obtain the final result.
Our multi-view synthesis pipeline is quite efficient since it involves no run-time training/optimization, only forward inferences by applying pre-trained models for occlusion analysis and multi-view synthesis.
Our method predicts camera trajectories that reveal occlusions effectively and produce consistent novel views, significantly improving 3D reconstruction over SOTA alternatives on the unseen GSO dataset. 
\end{abstract}
   
\section{Introduction}
\label{sec:intro}

\begin{figure*}[h]
    \centering
    \includegraphics[width=0.95\textwidth]{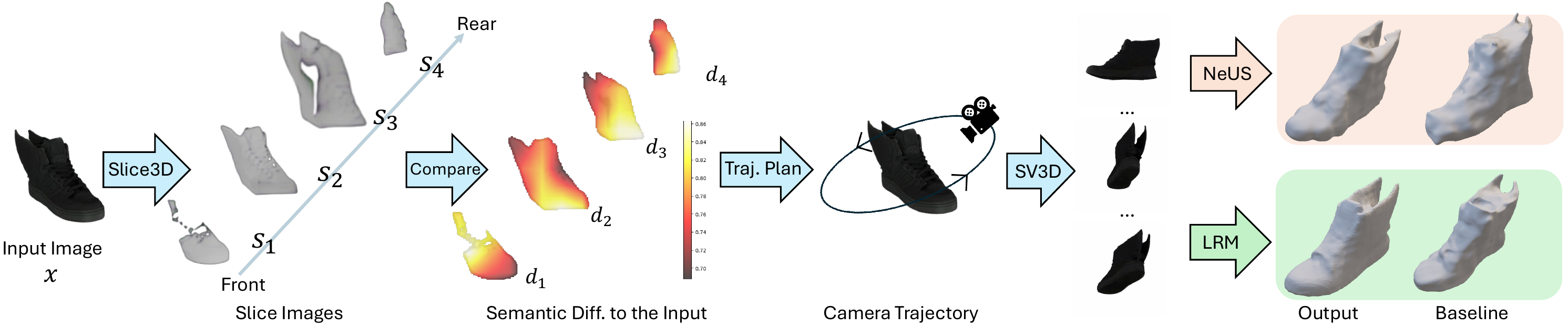}
\vspace{-1em}
\caption{Pipeline of our single-view reconstruction method, ACT-R, with adaptive camera trajectories (ACT). We first employ Slice3D~\cite{wang2024slice3d} to produce the slice images of the input object, with the slicing direction from the camera to the object center. Then we compute the semantic difference between the input and its slices by comparing their $512 \times 7 \times 7$ feature maps extracted from VGG16~\cite{simonyan2014very}. Each difference map $ d_i \in [0,1]^{7 \times 7}$ is up-scaled and overlaid onto slice images for beter visualization. Next, we identify the regions that have significant semantic differences (Sec.~\ref{sec:semantic_diff}), and plan the camera trajectories based on them (Sec.~\ref{sec:traj_plan}). Finally, we condition SV3D~\cite{voleti2024sv3d} on our planned trajectories, yielding a sequence of views, which can be fed into NeUS~\cite{wang2021neus} or InstantMesh (IM)~\cite{xu2024instantmesh} for multi-view 3D reconstruction.
}
\label{fig:pipeline}
\vspace{-1 em}
\end{figure*}

Single-view 3D reconstruction has been one of the most intensively studied problems in computer vision. One class of modern approaches directly generate or regress 3D representations of 
objects from input images~\cite{zhang2024clay,liu2024one,wu2024direct3d,xiang2024structured, hunyuan3d22025tencent, zhao2023michelangelo}, often resorting to 3D supervision which requires large 3D datasets for training. With significant advances in novel view synthesis~\cite{nerf,kerbl20233d},
the second line of popular approaches to single-view 3D reconstruction
first perform a multi-view synthesis, which typically generates an unordered set of views either {\em independently\/}~\cite{liu2023zero} or {\em simultaneously\/}~\cite{liu2023syncdreamer,long2024wonder3d,tang2024mvdiffusion++,shi2023zero123++} with {\em fixed camera setups\/}. This is followed by multi-view 3D reconstruction via differentiable rendering (e.g., NeUS~\cite{wang2021neus} and variants~\cite{long2022sparseneus,vora2023DiViNet}) so that the entire solution pipeline can avoid direct 3D supervision\footnote{We use the term ``direct 3D supervision'' in the context of 3D reconstruction to refer to models trained using paired images and ground-truth 3D models.}.

The main challenges to multi-view synthesis are twofold. First, the produced images should reveal {\em occluded\/} structures of the target 3D object that are hidden from the input image.
Second, the generated views must attain {\em 3D consistency\/} to ensure that a plausible and coherent 3D model 
can be reconstructed. Upon close examination of state-of-the-art multi-view synthesis methods, as well as single-view
reconstruction methods which combine such syntheses with direct 3D prediction~\cite{li2023instant3d,wang2024crm,tang2025lgm,liu2024meshformer}, we find that there is still much room for improvement.

In this paper, we introduce {\em adaptive view planning} to multi-view synthesis, so as to improve 
both occlusion revelation and 3D consistency for single-view 3D reconstruction. Instead of synthesizing an unordered set of views 
as in prior works, we generate a {\em sequence\/} of views, leveraging the inherent 
temporal consistency to enhance 3D coherence. More importantly, our view sequence is not constructed by a pre-determined setup~\cite{liu2023syncdreamer,long2024wonder3d,tang2024mvdiffusion++,melas20243d,yang2024viewfusion}. 
Instead, we compute an {\em adaptive camera trajectory\/} (ACT), which is specific to the target 3D object and its input view, to generate the sequence of novel views.
Note that several recent works have shown benefits of camera perturbation~\cite{voleti2024sv3d} and stochastic conditioning~\cite{watson2023novel} for novel view synthesis. Our approach takes these further with {\em judicious view planning}.

Given a single-view image of an object, we first search for an {\em orbit\/} of camera views to maximize the visibility of its 
occluded regions. Since occlusion prediction from a single image is ill-posed and searching over all orbits is intractable, we resort to a heuristic sampling of a manageable number of candidate orbits
and utilize a neural model to analyze occlusion revelation by the camera orbits. 

As our neural model, we employ Slice3D~\cite{wang2024slice3d}, a recent method for single-view 3D reconstruction which excels 
at recovering occluded 3D object structures. We apply a pre-trained Slice3D to predict a stack of images capturing parallel volumetric 
slices of the 3D object in the input image. We then rank the candidate camera orbits based on how well they reveal occluded regions 
of the 3D object, which can be localized over the slice images by examining semantic differences between them and the input image.

Once the best orbit is found, we feed it to a video diffusion model to generate a sequence of novel views that are adapting to the 3D 
object. The multi-view images obtained are finally passed to a 3D reconstruction model to obtain the final result. Note that these last two steps can employ a variety of state-of-the-art video diffusion and multi-view 3D reconstruction models. In our current work, we employ
Stable Video 3D (SV3D)~\cite{voleti2024sv3d} for the former, while for the latter, either NeUS~\cite{wang2021neus}, a well established method, or InstantMesh (IM)~\cite{xu2024instantmesh}, a more recent one based on large reconstruction models (LRMs). As a result, our entire solution pipeline (see Fig.~\ref{fig:pipeline}), which is coined ACT-R for using \underline{A}daptive \underline{C}amera \underline{T}rajectory for 
single-view 3D \underline{R}econstruction, does \emph{not} use direct 3D supervision since none of Slice3D, NeUS, or InstantMesh does.
Also, our multi-view generation is quite efficient since it involves no run-time training or optimization, only forward inferences by applying the pre-trained Slice3D and SV3D.

\begin{figure}[!t]
    \centering
    \includegraphics[width=\linewidth]{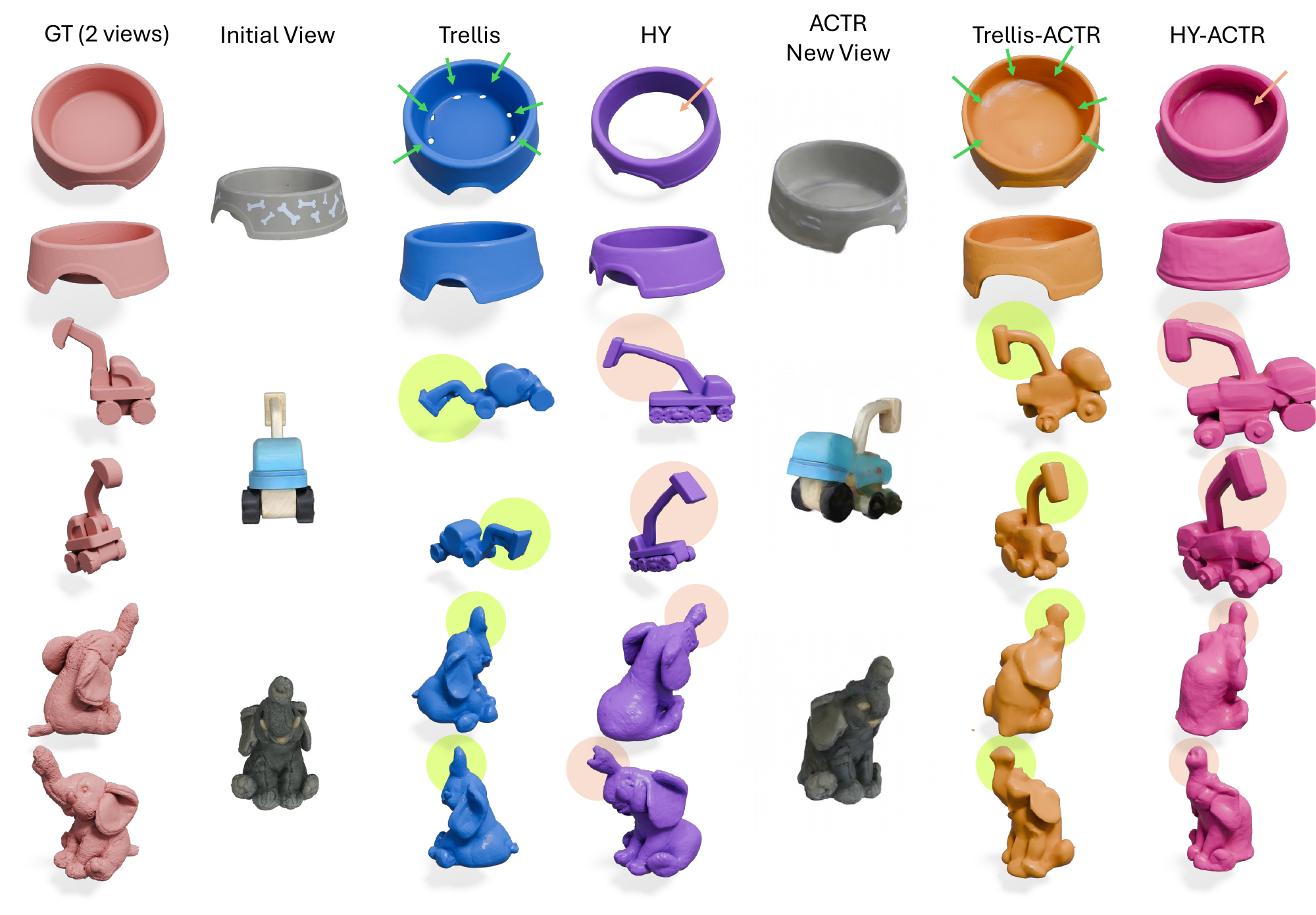}
\vspace{-0.3cm}
\caption{Initial single-view inputs to advanced methods such as Trellis and Hunyuan3D (HY) can be improved by one of the synthesized multi-view images from ACT-R, to improve reconstruction results, as highlighted in the colored regions.}
\vspace{-2 em}
\label{fig:ACTR-Trellis}
\end{figure}

Extensive experiments demonstrate that our method predicts camera trajectories tailored to each input example, effectively revealing occluded regions for higher-quality reconstruction. On the GSO benchmark~\cite{downs2022google}, ACT-R outperforms, both qualitatively and quantitatively, state-of-the-art methods based on multi-view syntheses from static~\cite{voleti2024sv3d}, fixed~\cite{tang2025lgm,long2024wonder3d}, or randomly perturbed~\cite{voleti2024sv3d} camera trajectories, all without direct 3D supervision, as well as a recent method, Craftsman~\cite{li2024craftsman}, which employs 3D supervision in the multi-view reconstruction step.
On the other hand, ACT, which is best positioned to boost multi-view 3D reconstruction {\em without\/} direct 3D supervision, cannot beat some of the latest LRMs such as CLAY~\cite{zhang2024clay}, Hunyuan3D~\cite{hunyuan3d22025tencent}, or Trellis~\cite{xiang2024structured}, which all use direct 3D supervision trained by closed-source (CLAY) or carefully curated (Trellis) large-scale 3D assets. Directly comparing ACT-R with these models can be unfair due to the different training setups. Nevertheless, as shown in Fig.~\ref{fig:ACTR-Trellis}, ACT can benefit these supervised LRMs by improving the input views.

\section{Related Work}
\label{sec:formatting}

As single-view 3D reconstruction has been extensively studied, we only focus on closely related works on video generation, multi-view synthesis, and view planning.


\subsection{Video Generation in 2D and for 3D}
Early works~\cite{ho2022video,singer2022make,ho2022imagen,ge2023preserve,bar2024lumiere,wu2023tune,zhang2023i2vgen,zhou2022magicvideo} extend image diffusion to video generation. Stable Video Diffusion (SVD)~\cite{blattmann2023stable} adapts latent diffusion methods~\cite{rombach2022high,blattmann2023align} to large-scale video datasets for temporally coherent generation.
A stream of work focuses on keyframe conditioning~\cite{girdhar2023emu, wang2024microcinema, li2023videogen, zeng2024make}, where initial frames are generated to anchor subsequent video synthesis, with latent-consistency networks ensuring temporal and appearance coherence. 
Training-free approaches~\cite{hong2023large,huang2024free} utilize Large Lanague Models (LLMs) for generation guidance.
Although video generation models typically lack explicit 3D representations, they can still achieve 3D consistency by producing temporally coherent videos. In our approach, we use video diffusion models as the backbone to generate view sequences.

The temporal coherency achieved by video generation models such as SVD~\cite{blattmann2023stable} can be utilized to enhance 3D consistency.
CameraCtrl~\cite{he2024cameractrl} trains a camera encoder upon a pre-trained T2V (Text-to-Video) model (such as~\cite{guo2023animatediff}) to allow precise and customizable camera pose control. ViewCrafter~\cite{yu2024viewcrafter} uses partial point clouds to render images as the condition of video diffusion models.
CamCo~\cite{xu2024camco} employs Plücker coordinates to control camera poses and leverages epipolar constraints to enhance 3D consistency in generated videos.
Finally, SV3D~\cite{voleti2024sv3d} fine-tunes a video generation model~\cite{blattmann2023stable} to create orbital videos around a 3D object given a camera trajectory. By adding random perturbations to the camera orbit, it can reveal structures that a standard orbit cannot. 
However, these randomized orbits do not reliably reveal additional occlusions. As more views are generated, the process slows down considerably.

\vspace{-5pt}

\subsection{Multi-view Synthesis from Single View}
Multi-view images are commonly adopted as an intermediate representation between single view and 3D, by which novel views are first synthesized from the input view, and then 3D entities are reconstructed via optimization ~\cite{long2024wonder3d, voleti2024sv3d} or feed-forward networks~\cite{li2023instant3d,xu2024instantmesh,tang2025lgm}.  Zero-1-to-3~\cite{liu2023zero} fine-tunes Stable Diffusion (SD)~\cite{rombach2022sd} to generate novel views conditioned on camera poses.
One-2-3-45~\cite{liu2024onev1} and its upgraded version~\cite{liu2024one} combines such multi-view image generator with a feed-forward network, achieving reconstruction speeds as few as 45 seconds. Such a pipeline is later improved in Instant3D~\cite{li2023instant3d}, InstantMesh~\cite{xu2024instantmesh}, and LGM~\cite{tang2025lgm} for faster and higher-quality inference by leveraging large reconstruction models (LRM).
SyncDreamer~\cite{liu2023syncdreamer} enhances 3D consistency through cross-attention mechanisms between different views.
Wonder3D~\cite{long2024wonder3d} produces depth and normal maps alongside with the RGB novel views to further support accurate 3D reconstruction.
MVDiffusion~\cite{tang2023mvdiffusionenablingholisticmultiview} and MVDiffusion++~\cite{tang2024mvdiffusion++} utilize cross-view attention to improve multi-view consistency for generating panoramas and 3D structures.

Generating multiple views from a single image in one step presents significant consistency challenges, as establishing correspondences between substantially different viewpoints remains difficult.
To address this issue, 3DiM~\cite{watson2023novel} generates novel views in an auto-regressive manner, selecting a previously generated view as a condition for producing each subsequent view during the denoising process.
ViewFusion~\cite{yang2024viewfusion} builds upon Zero-1-to-3~\cite{liu2023zero} to generate novel views using a similar auto-regressive approach as 3DiM~\cite{watson2023novel}.
IM-3D~\cite{melas20243d} employs a video generation model~\cite{girdhar2023emu} to create novel views that are then processed by 3D G-Splat~\cite{kerbl20233d} for 3D reconstruction. This approach can be iteratively refined by feeding rendered objects back into the video diffusion model.
However, these methods (3DiM, ViewFusion, IM-3D) rely on either random camera poses or predefined trajectories for novel view generation, without considering the specific structural properties of the objects being modeled.


\vspace{-5pt}

\subsection{View and Path Planning}

Path planning has applications in navigation, scanning, and even computational fabrication~\cite{galceran2013survey,karur2021survey,roberts2016generating,roberts2017submodular,liu2022learning,VDACAli}.
In particular, Next-Best-View (NBV) planning addresses the fundamental challenge of determining an optimal sequence of camera positions to maximize information gain during scene or object inspection ~\cite{connolly1985determination, massios1998best, vasquez2014volumetric}. Two popular traditional approaches to this problem include voxel-space methods that optimize coverage metrics ~\cite{connolly1985determination, massios1998best, vasquez2014volumetric, maldonado2016next}, and surface-based methods that analyze boundary characteristics to determine optimal viewpoints ~\cite{pito1999solution, kriegel2011surface, chen2005vision}.
Recent advances in deep learning have transformed the NBV paradigm, introducing reinforcement learning~\cite{peralta2020next,chen2024gennbv}, reconstructability predictor~\cite {liu2022learning} and uncertainty evaluation framework~\cite{jin2023neu} to the scope. While traditional NBVs assume the availability of a complete 3D reference model, our work addresses a more challenging scenario where only a single image serves as input. This constraint fundamentally shifts the problem from pure coverage optimization to view prediction based on limited initial information, requiring novel strategies for trajectory planning.



\section{Method}
\label{sec:method}

Given a single image $x \in \mathbb{R}^{3 \times H \times W }$ of an object and a video generation model $\mathcal{G}$ conditioned on a camera trajectory (e.g., SV3D~\cite{voleti2024sv3d}), our goal is to determine an adaptive camera pose trajectory $(\pi_{1}, \pi_{2}, ..., \pi_{N})$ that adapts to the object in its input view. It should condition \( \mathcal{G} \) to generate a sequence of views that better reveals the geometry of the target object compared to a fixed generic trajectory, and leads to a more accurate 3D reconstruction.

Similar to SV3D~\cite{voleti2024sv3d}, we assume that the camera always looks at the center of an object (origin of the world), and the distance between the camera and the center remains unchanged, so any viewpoint can be specified by only two parameters:
$\pi_{i} = (e_i, a_i)$, where $e_i, a_i$ are the elevation and azimuth angles.

When designing an adaptive trajectory in contrast with a generic one, we aim to avoid increasing its length (i.e., the number of views), as doing so would complicate both the multi-view generation and subsequent 3D reconstruction. To maintain consistency, we fix the number of generated views, $N$, as $21$ and use a single closed orbit for the camera trajectory,
the same as SV3D (u)~\cite{voleti2024sv3d} for a fair comparison as SV3D serves as our closest baseline.

\subsection{Overview}
\label{subsec:overview}

Fig.~\ref{fig:pipeline} shows our pipeline. To guide camera trajectory generation, we first identify occluded regions from the input view. To this end, we employ Slice3D~\cite{wang2024slice3d} to produce object slice images from the front to the rear of the object. These slices allow us to construct a coarse representation by voxelizing each slice and stacking them up.
When comparing each slice image to the input image, significant semantic differences in certain areas often suggest that these regions in the slice are occluded from the input view. Therefore, we compute semantic difference maps between each slice image and input image by leveraging VGG16 features and incorporating this information into the voxels, forming a series of spatially-aware 3D semantic difference blocks.

From these blocks, we plan a camera trajectory aiming to maximize the visibility of the occluded regions, with each block weighted by its semantic difference.
This optimized trajectory is fed into SV3D~\cite{voleti2024sv3d} to generate a sequence of novel views. The generated novel views can be used in a plug-and-play manner for any 3D reconstruction pipeline that takes novel views as intermediate representations. We used NeUS~\cite{wang2021neus} and large reconstruction models trained from InstantMesh~\cite{xu2024instantmesh} as our two alternative 3D reconstruction backbones. 

\subsection{Building Semantic Difference Blocks}
\label{sec:semantic_diff}

\begin{figure}[!t]
    \centering
    \includegraphics[width=0.99\columnwidth]{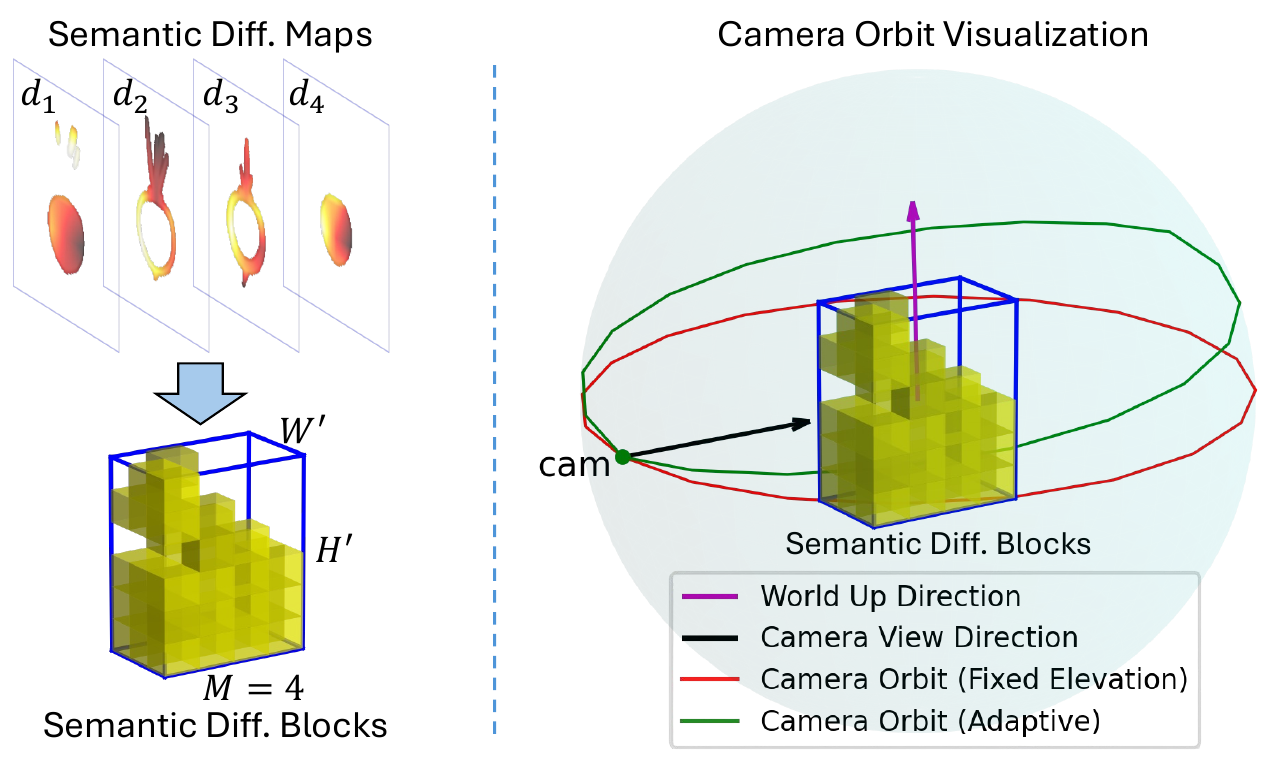}
\vspace{-0.5em}
\caption{Illustration of camera trajectory planning. Left: Transforming the semantic difference maps into 3D blocks, where lighter yellow indicates greater differences. Right: Visualization of different camera orbits. Red: fixed elevation; Green: variable elevations that capture greater visibility. ``diff'' and ``cam'' denote difference and camera, respectively.}
\label{fig:cmr_planning}
\end{figure}


As illustrated in the first step in Fig.~\ref{fig:pipeline}, Slice3D produces in total $M$ slice images $\{s_1, s_2, \dots, s_M\}$ from the input $x$.
We quantify the semantic differences between slice image $s_i$ and input $x$ by comparing the features from the final pooling layer of VGG16~\cite{simonyan2014very}. Specifically, we compute:
$d_i = \langle \phi(x), \phi(s_i) \rangle,$
where $\phi(\cdot) \in \mathbb{R}^{512 \times 7 \times 7}$ represents the VGG16 feature maps for a given input, with spatial resolution $7 \times 7$ and feature dimension $512$ per position; $1 \leq i \leq M$; $\langle \cdot, \cdot \rangle$ denotes the cosine similarity between the 512-dimensional feature vectors at each corresponding spatial position from the two inputs, resulting in $d_i \in \mathbb{R}^{7 \times 7}$.

As shown in Fig.~\ref{fig:cmr_planning} (left), we transform the semantic difference maps $\{d_1, ..., d_M\}$ into 3D blocks (shown in yellow), approximating the object shape and indicating semantic differences to the input view.
Specifically, we first identify the 2D mask of the object in the input image, and apply this mask to crop each map $d_i$ into $d^{\prime}_{i} \in \mathbb{R}^{H' \times W'}$ ($H' \leq 7$, $W' \leq 7$). 
Next, we estimate the object's 3D bounding box in the camera coordinate system (detailed in the supp. material) and divide this space into $M \times H' \times W'$ blocks. Each block at position $(i,j,k)$ corresponds to the element $d^{\prime}_{ijk}$, where $1 \leq i \leq M$, $1 \leq j \leq H'$ and $1 \leq k \leq W'$. 

We retain a block only when its corresponding slice pixel $s_{ijk}$ is not empty. The retained block is associated with the value of $d^{\prime}_{ijk}$ indicating the semantic difference of that region compared to the input view.
This process yields a coarse 3D representation of the input object, highlighting which parts were obstructed from the input view and require further observation from subsequent views.

Compared with more recent feature encoders such as DINOv2~\cite{oquab2023dinov2} and CLIP~\cite{radford2021learning}, we empirically find that VGG16~\cite{simonyan2014very} works better in extracting semantic difference maps between slices and input images. Please see the supplementary materials for detailed ablation studies.

\subsection{Camera Trajectory Planning}
\label{sec:traj_plan}

To accurately model the spatial relationship between the camera and the reconstructed object, we estimate:

\begin{enumerate}
\item $r$: distance from the camera to the object center, with the camera trajectory as an orbit on a sphere with radius $r$.
\vspace{3pt}
\item $\alpha$: the elevation over the object in the input image.
\end{enumerate}


The camera is initially positioned at the input view, i.e., $(a_0, e_0) = (0, \alpha)$. The generic camera trajectory from SV3D (u)~\cite{voleti2024sv3d} uses a fixed elevation $\alpha$, shown as the red orbit in Fig.~\ref{fig:cmr_planning} (right). Here $r$ and $\alpha$ are used to calculate the world-to-camera coordinate transform since SV3D~\cite{voleti2024sv3d} operates in a world frame while we use a camera coordinate system.

The reconstructed coordinate system is shown in Fig.~\ref{fig:cmr_planning}. To maximize the observation of occluded regions, an ideal trajectory should aim to cover as many blocks as possible, prioritizing those blocks with greater semantic differences (denoted by lighter-coloured blocks). With the camera's known field of view (FOV) $\theta$, we can determine the block visibilities from any given position.

Since the search space for camera trajectories is infinite, we discretize the camera's movement to facilitate the search. Specifically, we begin by sampling the azimuth steps at a constant interval of $18^\circ$ (calculated as $360^\circ/20$) 
between each frame 
from $0^\circ$ to $360^\circ$, which creates a closed-loop trajectory in terms of azimuth.

Next, we define the elevation angle changes at each step using the set $ \{\pm5^\circ, \pm4^\circ, \pm3^\circ, \pm2^\circ, \pm1^\circ, 0^\circ\}$. We limit per-step elevation changes to within $5^\circ$, as larger step sizes challenges frame-consistent video generation. Each orbit is divided into four segments based on azimuth angles. Within each segment, the elevation angle increments at a constant rate, with the step size selected from the set.
To ensure a closed orbit, we enforce the total variation in elevation to be zero by mirroring and negating the elevation change in the second segment onto the third and the first onto the fourth. 
 This approach results in a total of $11 \times 11 = 121$ candidate trajectories, which we denote as the set $\Pi$.


We choose the path $\pi^*$ that maximizes the weighted visibility of difference blocks. For each camera position, we determine which blocks are visible, and our objective is to optimise culmulative visibility weight across all time steps:

\begin{equation}
\label{equ:vis_score}
    \pi^* = \operatorname*{arg\,max}_{\pi \in \Pi} \sum_{t=1}^{N} \sum_{\substack{(i,j,k) \\ \in \psi(\pi(t))}} d^{\prime}_{ijk},
\end{equation}
where $\psi(\pi(t))$ denotes the set of all visible blocks under camera $\pi(t)$. More details about the $\psi(\cdot)$ formulation and orbital camera trajectory justification can be found in the supplementary material.

\subsection{View Generation and 3D Reconstruction}

We fed the camera trajectory $\pi^*$ to SV3D~\cite{voleti2024sv3d} to generate a video containing a sequence of novel views.
Since the video could suffer from significant artifacts due to stochasticity in the generative models, we apply view-consistency as the primary metric to filter out low-quality results and regenerate with alternative random seeds when necessary. The final view sequence can be integrated into any 3D reconstruction pipeline that accepts posed-multiview-images as input. We demonstrate ACT-R's flexibility by reconstruction meshes through two different approaches: volumetric rendering with NeUS~\cite{wang2021neus} using all 21 generated images, and processing through LRM~\cite{xu2024instantmesh} by feeding 6 uniformly sampled key frames from the 21 views.
\section{Experiments}
\label{sec:exp}

\mypara{Implementation details}
We used \texttt{rembg}~\cite{qin2020u2} to remove the background of input images.
The number of slices $M$ is set to 4. The FOV $\theta$ of the camera is set to 33.8.
SV3D produces a video consisting of 21 frames, with each frame at a resolution of $576 \times 576$ pixels.
For volumetric rendering based reconstruction, we chose NeUS~\cite{wang2021neus} as our reconstruction method because it is well-established and continues to be used in recent state-of-the-art methods (e.g., Wonder3D~\cite{long2024wonder3d}). Our 3D reconstruction performance could potentially be enhanced with more recent 3D reconstruction methods, such as those in~\cite{huang20242d,li2023neuralangelo}.
We train the network of NeUS for 10k steps for each shape, which takes around 15 minutes in an NVIDIA 3090 GPU. For LRM based reconstruction,  we uniformly sampled 6 views from the view sequence used pretrained checkpoint from InstantMesh~\cite{xu2024instantmesh} as it takes images from arbitrary camera pose.

\vspace{2pt}

\mypara{Datasets}
We conduct evaluation on GSO~\cite{downs2022google}, a widely-used benchmark for novel view synthesis and 3D reconstruction which includes about 1K common household objects that were 3D scanned and represented as meshes.

\vspace{2pt}

\begin{figure*}[htbp!]
    \centering
    \includegraphics[width=0.95\textwidth]{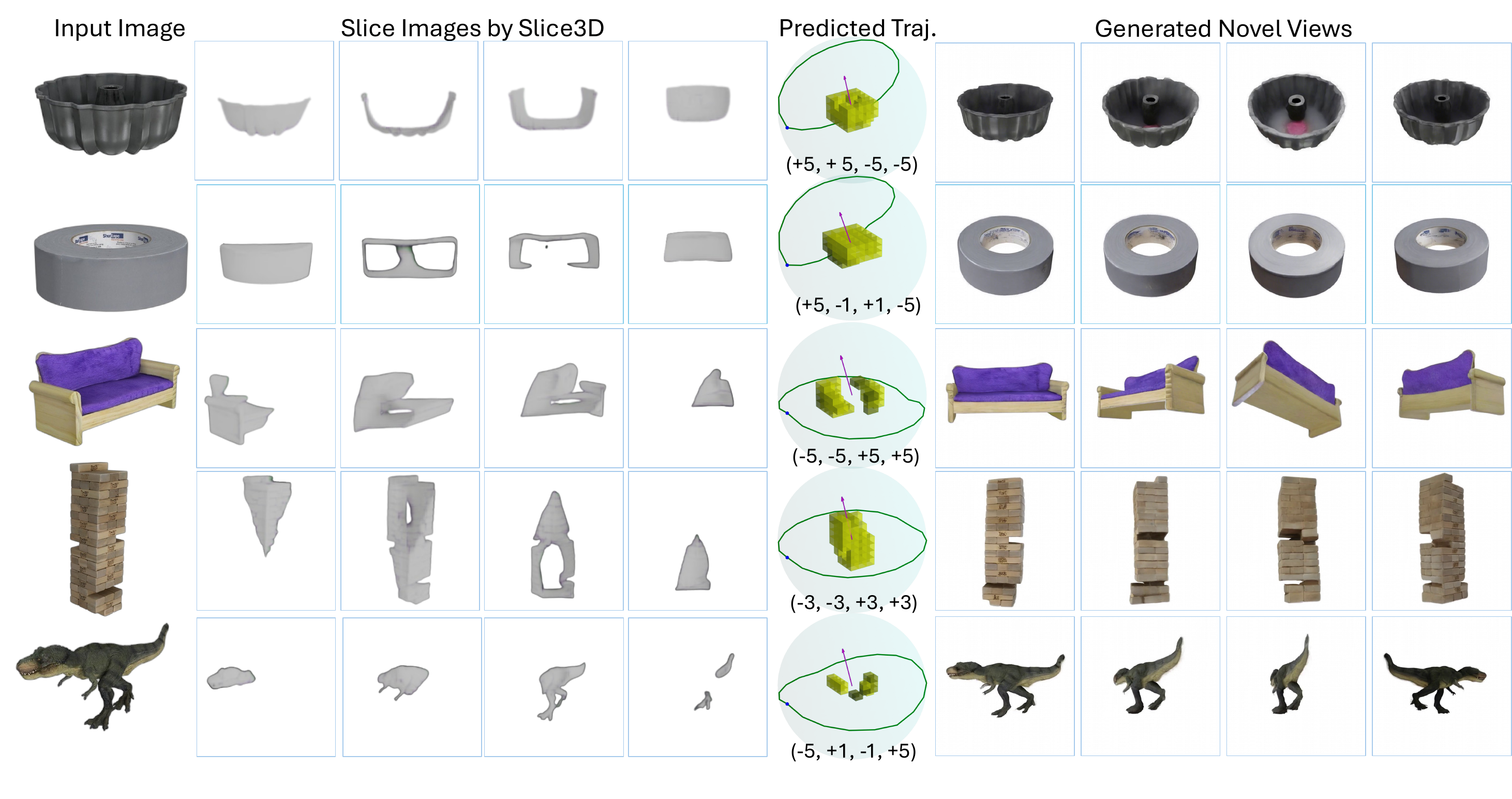}
\vspace{-1em}
\caption{Visualization of slice images predicted by Slice3D~\cite{wang2024slice3d} and planned trajectories (shown in green). Purple arrows indicate the world's up direction. 
The 3D blocks (i.e., semantics difference blocks) roughly represent the input shape, using colors to highlight their semantic differences from the input view. The numbers in brackets show the elevation changes in each segment of the orbit.
}
\label{fig:vis_of_slices_trajs}
\end{figure*}

\mypara{Evaluation Metrics}
For 3D metrics, we use Chamfer $\mathcal{L}_{2}$ (CD), Hausdorff (HD), and Light Field distances (LFD)~\cite{chen2003visual}, as well as F-score\%1 (F1)~\cite{tatarchenko2019single}, to evaluate reconstruction results.
In addition, we also report 2D metrics such as PSNR, SSIM~\cite{wang2004image}, and LPIPS~\cite{zhang2018unreasonable} on 12 rendered views of the reconstructed meshes. Since no single metric is entirely informative, we provide a comprehensive evaluation to cover both object- and image-space, as well as distortion and visual similarity aspects of the 3D reconstruction.

\vspace{-3pt}

\subsection{Qualitative Visual Comparisons}

We compare ACT-R to representative SOTA approaches for single-view 3D reconstruction, including Wonder3D~\cite{long2024wonder3d}, SV3D(u)~\cite{voleti2024sv3d} for volumetric-rendering-based reconstruction, LGM~\cite{tang2025lgm} based on LRMs, and CraftsMan~\cite{li2024craftsman} as an image-to-3D method with direct 3D supervision. We further present three ablated experiment setups:
\begin{enumerate}
    \item Random trajectory: Generate novel views with a random orbital trajectory by using random elevation increments, with the mesh reconstructed by NeUS.
    \item $\text{SV3D}_{\text{IM}}$: LRM with 6 views sampled uniformly from generic camera trajectories with constant elevation.
\end{enumerate}  

\begin{figure*}[!t]
    \centering
    \includegraphics[width=\textwidth]{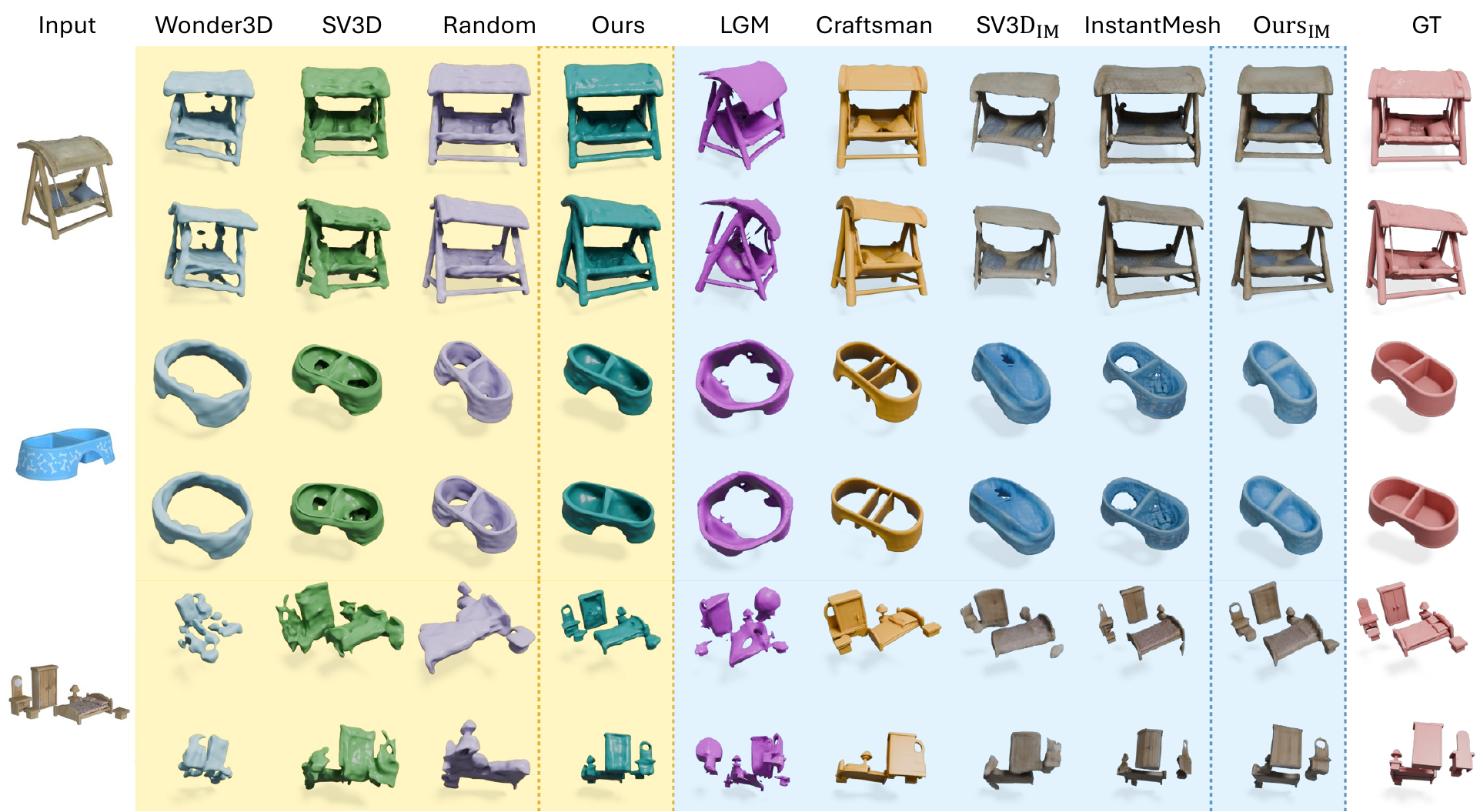}
\caption{Qualitative visual comparisons between single-view 3D reconstruction methods on the GSO dataset. Please zoom in for a closer inspection. Meshes reconstructed with NeUS and LRM are in yellow and blue blocks, respectively. Our adaptive camera trajectories can be integrated into both reconstruction frameworks. From these examples, it is evident that our method can capture geometric and structural details better, especially over concavities and occluded regions, and it has less missing parts and geometric artifacts. For example, our method is the only one, whether with NeuS or LRM, that can faithfully reconstruct the inside region of the dog bowl (second example).}
\vspace{-0.2cm}
\label{fig:comp_qual_gso}
\end{figure*}

Qualitative results in ~\Cref{fig:comp_qual_gso} show that
different reconstruction backbones offer distinct advantages, while also expose their limitations. NeUS produces more faithful reconstructions that closely match the input, yet suffers from blobby surface artifacts. Effectively leveraging 3D priors, LRMs produce smoother surfaces, but respect less the details from the input image. Despite these inherent backbone-specific characteristics, ACT-R's adaptive trajectories {\em consistently\/} improve reconstruction quality, particularly over occluded regions. Our approach more faithfully reconstructs the solid bowl interiors (rows 2) that most competing methods overlooked due to occlusion in the input image and provides enhanced spatial awareness between objects (row 3). See supplementary material for more results.


\begin{figure}
    \centering
    \includegraphics[width=0.95\columnwidth]{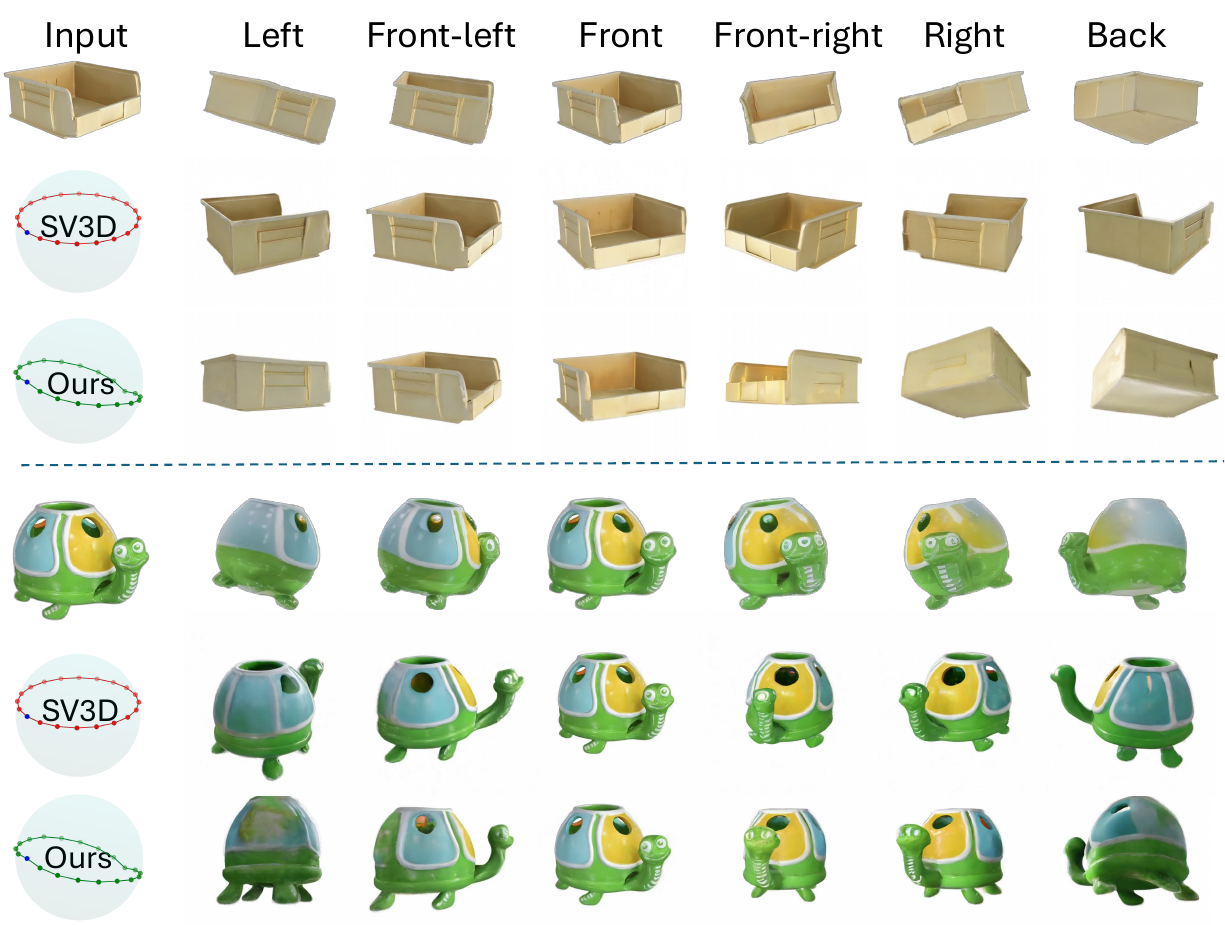}
\vspace{-0.5em}
\caption{Generated views of Wonder3D~\cite{long2024wonder3d} (1st row), SV3D~\cite{voleti2024sv3d} (2nd row) and our method (3rd row). The red and green orbits show the orbits of SV3D and our method, respectively.
\vspace{-1.5 em}}
\label{fig:vis_gen_views}
\end{figure}

Fig.~\ref{fig:vis_gen_views} compares the generated views of Wonder3D~\cite{long2024wonder3d} and SV3D~\cite{voleti2024sv3d} to ours. Compared to SV3D, Wonder3D offers a wider range of visibility (e.g., bottom views of the objects) by varying both camera azimuths and elevations. However, it tends to suffer from multi-view inconsistency, as seen in the heavily distorted back view. SV3D maintains better consistency but cannot observe the back of the object, leading to inadequate information during 3D reconstruction. In contrast, our method captures the bottom of the object and achieves view consistency.

\subsection{Quantitative Comparisons}

Since the reconstructed 3D meshes from different methods exhibit different poses, we first normalize the output mesh and provide an initial transform to align it with the GT mesh, then optionally used Iterative Closest Point (ICP) to further tune the mesh poses, whichever leads to better quantitative measures for the results. 
 
We report quantitative comparison results for all 1,030 objects from the GSO dataset in Table~\ref{tab:comp_quan_gso}. Our method demonstrates clear advantages over the other methods across all metrics, especially in F1~\cite{tatarchenko2019single}.
The improvements on 2D metrics indicate that our method can generate objects that are aligned with the GT in pixel space. Overall, our method exhibits robustness in accurately modeling object geometry while preserving visual realism in image-space projections.
Its superior performance on GSO further underscores its strong generalizability across diverse object categories, showcasing its capability in handling a wide range of shapes, sizes, and appearances effectively. 
 

    

\begin{table}[t]
    \centering
    \resizebox{\columnwidth}{!}{
    \begin{tabular}{cccccccc}
     \toprule
     Method               & CD$\downarrow$  & F1$\uparrow$ & HD$\downarrow$ & PSNR$\uparrow$ & SSIM$\uparrow$ & LPIPS$\downarrow$ & LFD $\downarrow$ \\ 
     \midrule
     \addlinespace[4pt] 
     Wonder3D~\cite{long2024wonder3d}            &  5.17             & 2.66  & 18.9 & 15.8 & 8.17 &  17.9 & 1.33 \\
     
     SV3D(u)~\cite{voleti2024sv3d}      & 4.93 & 2.79 & 18.4 & 15.9 & 8.12 & 17.6 & 1.30 \\
     
     LGM~\cite{tang2025lgm} & 7.78 & 1.53 & 28.9 & 13.3 & 7.56 & 24.1 & 1.79 \\
     
     Craftsman~\cite{li2024craftsman} & 6.37 & 2.81 & 20.8 & 15.7 & 8.08 & 17.4 & 1.42 \\
     
     $\text{SV3D}_\text{IM}$ & 5.15 & 2.64 & 18.0 & 16.4 & \cellcolor{secondplace!30}8.35 & 16.1 & 1.17 \\
     
     Random Traj. & \cellcolor{thirdplace!30}4.79 & \cellcolor{thirdplace!30}3.15 & \cellcolor{secondplace!30}17.2 & \cellcolor{thirdplace!30}16.8 & \cellcolor{thirdplace!30}8.32 & \cellcolor{thirdplace!30}15.5 & \cellcolor{secondplace!30}1.04 \\
     
     $\text{Ours}_\text{IM}$ & \cellcolor{secondplace!30}4.51 & \cellcolor{secondplace!30}3.41 & \cellcolor{firstplace!30}\textbf{16.2} & \cellcolor{firstplace!30}\textbf{17.4} & \cellcolor{firstplace!30}\textbf{8.49} & \cellcolor{firstplace!30}\textbf{14.0} & \cellcolor{thirdplace!30}1.05 \\
     
     Ours & \cellcolor{firstplace!30}\textbf{4.47} & \cellcolor{firstplace!30}\textbf{3.78} & \cellcolor{thirdplace!30}17.5 & \cellcolor{secondplace!30}17.1 & \cellcolor{secondplace!30}8.35 & \cellcolor{secondplace!30}15.1 & \cellcolor{firstplace!30}\textbf{1.00} \\
     \bottomrule
    \end{tabular}
    }
    \caption{Comparison of 3D reconstruction results. Cell colors indicate ranking: \textcolor{firstplace!80}{green} (1st), \textcolor{secondplace!80}{blue} (2nd), and \textcolor{thirdplace!80}{amber} (3rd) for each metric. Lower is better for CD, HD, LPIPS, and LFD. Higher is better for F1, PSNR, and SSIM.
   }
    \label{tab:comp_quan_gso}
    \vspace{-1em}
\end{table}

We further compared the coverage metric resulting from different trajectories. Quantitative numbers suggest that our adaptive trajectory achieves a better coverage than static or random trajectories, please see the supplementary material for more details.

\vspace{-3pt}



\subsection{Visualizations of Trajectories}

Fig.~\ref{fig:vis_of_slices_trajs} shows our predicted slice images and trajectories.
An naive view planning would negate the initial elevation angle for an even span, raising the camera for negative angles and lowering it for positive ones. In contrast, we base our trajectory prediction on a deeper understanding of object structures and occlusions.

For the Bundt cake pan, even with a positive elevation for the input view, we still raise the camera to reveal the inner tube. Although slice3D~\cite{wang2024slice3d} failed to predict the inner tube in the sliced image, it still suggests that inner regions require additional attention. 
For the tape example, the camera highest elevation is lower that the pan example, since it has already fully observed the hole in the middle. This proves that our method is adaptive to each single object.
For the sneaker example, our trajectory can better observe the pair of wings so that they will not occlude with each other in most of the frames. For the sofa, we lower the camera to fully capture its bottom. For the Jenga blocks, we planned a trajectory that can better observe the concave areas.

\vspace{-3pt}

\subsection{Ablation Studies}
\begin{figure}
    \centering
    \includegraphics[width=0.95\columnwidth]{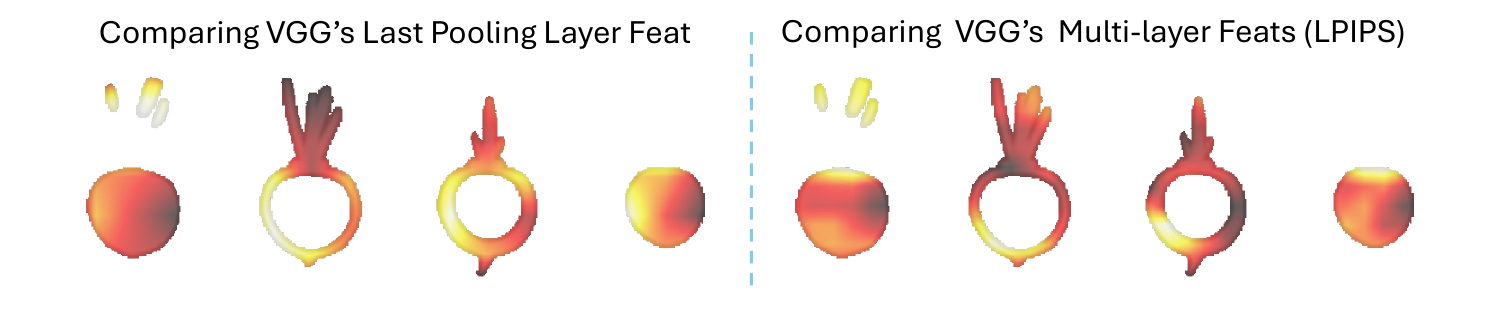}
\vspace{-0.5em}
\caption{Computing semantic difference maps $\{d_i\}$ using different metrics. Brighter color indicates higher semantic difference.}
\label{fig:compute_sm_diff}
\vspace{-1em}
\end{figure}

\mypara{Computation of semantic differences} The computation of semantic difference maps $\{d_i\}$ plays an critical role in our task. From the comparison between Ours and random trajectory result, it is evident that our current VGG features is guiding the camera in a meaningful way. Aside from VGG16, we also tried perceptual loss (LPIPS)~\cite{zhang2018unreasonable} to compute the semantic differences by comparing multi-level features from VGG, rather than the last-pooling layer features we currently employ. As shown in Fig.~\ref{fig:compute_sm_diff}, it appears that high-frequency (low-layer) features do not aid in locating the occluded regions. This is particularly evident in the third slice images, where LPIPS generates a dark map that barely highlights any differences from the input view. 

\vspace{2pt}

\begin{figure}[!t]
    \centering
    \includegraphics[width=0.95\columnwidth]{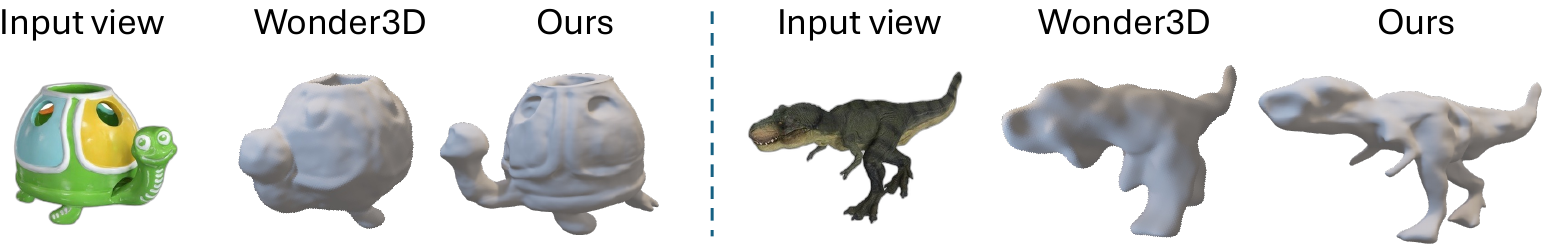}
\vspace{-0.5em}
\caption{Reconstruction from 6 generated views.}
\label{fig:rec_from_6_views}
\vspace{-1 em}
\end{figure}

\mypara{Reconstruction from 6 views} Since Wonder3D only generates 6 views to obtain a 3D mesh, we also test our method to only rely on 6 frames from our generated videos that have nearly the same azimuths as Wonder3D. The results in Fig.~\ref{fig:rec_from_6_views} indicate that the quality of the views is more important than their quantity.

\vspace{2pt}

\mypara{Robustness to camera pose estimation} Since our trajectory planning operates on a coarse 3D representation, i.e., the blocks, it is robust against errors from estimating the camera parameters and the bounding boxes.
More details on how camera estimation affects our view planning can be found in the supplementary material.

\section{Conclusion, Limitation, and Future Work}
\label{sec:future}
We propose adaptive view planning for synthesizing novel views from a single image. Our key insight is that slice images from Slice3D~\cite{wang2024slice3d} can effectively reveal occluded structures from the input view and guide the camera's movement. By leveraging the capabilities of modern video generation models, the generated novel views along our planned trajectory tend to improve the visibility of occluded structures while maintaining multi-view consistency and the overall frame budget.
Interestingly, this reveals how multi-slice and multi-view can work together to complement each other, where the former provides geometric insights on where the self-occlusion may be present while the later offers signals on the visible geometries. Combining the two, potential invisible regions are better exposed for improved single-view 3D reconstruction. In the end, our method ACT-R has been shown to outperform state-of-the-art reconstruction alternatives which take the single-to-multi view route without resorting to direct 3D supervision.

\vspace{2pt}

\mypara{Limitations}
Most limitations in our current implementation are inherited from SV3D~\cite{voleti2024sv3d}, e.g., limited camera DoFs as only azimuth and elevation are changeable, and quality of the generated videos. 
Nonetheless, our view planning is applicable to other video generation models such as ViewCrafter~\cite{yu2024viewcrafter}, VD3D~\cite{bahmani2024vd3d}, and CameraCtrl~\cite{he2024cameractrl} with even higher camera DoFs. 
ACT-R's performance is also limited by the accuracy of the estimated semantic differences between slice and input images. It is certainly possible to explore alternative occlusion-aware reasoning for view planning, or enlarge the scale of 3D data exposed to Slice3D --- the version we employed for trajectory planning was trained on only 5\% of the 880K Objaverse 3D dataset. 


\vspace{2pt}

\mypara{Adaptive vs.~fixed camera trajectories}
While we firmly believe in the merits of adaptive view trajectories for occlusion revelation, view synthesis from such trajectories (e.g., via SV3D~\cite{voleti2024sv3d}) is more difficult compared to one that only works with fixed cameras, e.g., Zero123++~\cite{shi2023zero123++}. Even when trained on the same 3D dataset, SV3D is clearly outdone by Zero123++ in terms of the quality of the synthesized multi-view images. As a result, our method, which feeds adaptive trajectories to SV3D, cannot quantitatively beat InstantMesh~\cite{li2023instant3d}, which employs Zero123++, across most metrics. Qualitatively however, we consistently observe that our method can outperform InstantMesh when the 3D objects have significant concavities or occlusions; see the two bowls in the fourth blue column in Fig.~\ref{fig:comp_qual_gso}. We are motivated to resolve the above discrepancy by improving camera-adaptive view synthesis.


\vspace{2pt}

\mypara{LRMs with direct 3D supervision}
The most successful single-view reconstruction models of late have predominantly been LRMs~\cite{zhang2024clay,hunyuan3d22025tencent,xiang2024structured} with direct image-to-3D and 3D supervision using large-scale 3D datasets. The strong 3D priors learned by these models appears to be diminishing the importance of multi-view synthesis. As we show in the supplementary material, multi-view Trellis does not outperform its single-view counterpart. 
This is also due in part to the blurriness of the synthesized multi-view images, regardless of the camera trajectories. 
That being said, hallucinations still abound either due to severe occlusions in the input view (see Fig.~\ref{fig:ACTR-Trellis}) or erroneous priors (e.g., symmetries by CLAY~\cite{zhang2024clay}), while potential overfitting to 3D training data remains a valid concern.

In future work, a more in-depth investigation into how to best integrate adaptive view planning into direct image-to-3D large reconstruction models (LRMs) is worth conducting. We would also like to explore other applications of ACT, e.g., for robotics and autoscanning~\cite{wu2014autoscanning}.

{
    \small
    \bibliographystyle{ieeenat_fullname}
    \bibliography{main}

\begin{thebibliography}{81}
\providecommand{\natexlab}[1]{#1}
\providecommand{\url}[1]{\texttt{#1}}
\expandafter\ifx\csname urlstyle\endcsname\relax
  \providecommand{\doi}[1]{doi: #1}\else
  \providecommand{\doi}{doi: \begingroup \urlstyle{rm}\Url}\fi

\bibitem[Bahmani et~al.(2024)Bahmani, Skorokhodov, Siarohin, Menapace, Qian, Vasilkovsky, Lee, Wang, Zou, Tagliasacchi, et~al.]{bahmani2024vd3d}
Sherwin Bahmani, Ivan Skorokhodov, Aliaksandr Siarohin, Willi Menapace, Guocheng Qian, Michael Vasilkovsky, Hsin-Ying Lee, Chaoyang Wang, Jiaxu Zou, Andrea Tagliasacchi, et~al.
\newblock Vd3d: Taming large video diffusion transformers for 3d camera control.
\newblock \emph{arXiv preprint arXiv:2407.12781}, 2024.

\bibitem[Bar-Tal et~al.(2024)Bar-Tal, Chefer, Tov, Herrmann, Paiss, Zada, Ephrat, Hur, Liu, Raj, et~al.]{bar2024lumiere}
Omer Bar-Tal, Hila Chefer, Omer Tov, Charles Herrmann, Roni Paiss, Shiran Zada, Ariel Ephrat, Junhwa Hur, Guanghui Liu, Amit Raj, et~al.
\newblock Lumiere: A space-time diffusion model for video generation.
\newblock \emph{arXiv preprint arXiv:2401.12945}, 2024.

\bibitem[Blattmann et~al.(2023{\natexlab{a}})Blattmann, Dockhorn, Kulal, Mendelevitch, Kilian, Lorenz, Levi, English, Voleti, Letts, et~al.]{blattmann2023stable}
Andreas Blattmann, Tim Dockhorn, Sumith Kulal, Daniel Mendelevitch, Maciej Kilian, Dominik Lorenz, Yam Levi, Zion English, Vikram Voleti, Adam Letts, et~al.
\newblock Stable video diffusion: Scaling latent video diffusion models to large datasets.
\newblock \emph{arXiv preprint arXiv:2311.15127}, 2023{\natexlab{a}}.

\bibitem[Blattmann et~al.(2023{\natexlab{b}})Blattmann, Rombach, Ling, Dockhorn, Kim, Fidler, and Kreis]{blattmann2023align}
Andreas Blattmann, Robin Rombach, Huan Ling, Tim Dockhorn, Seung~Wook Kim, Sanja Fidler, and Karsten Kreis.
\newblock Align your latents: High-resolution video synthesis with latent diffusion models.
\newblock In \emph{Proceedings of the IEEE/CVF Conference on Computer Vision and Pattern Recognition}, pages 22563--22575, 2023{\natexlab{b}}.

\bibitem[Chen et~al.(2003)Chen, Tian, Shen, and Ouhyoung]{chen2003visual}
Ding-Yun Chen, Xiao-Pei Tian, Yu-Te Shen, and Ming Ouhyoung.
\newblock On visual similarity based 3d model retrieval.
\newblock In \emph{Computer graphics forum}, pages 223--232. Wiley Online Library, 2003.

\bibitem[Chen and Li(2005)]{chen2005vision}
SY Chen and YF Li.
\newblock Vision sensor planning for 3-d model acquisition.
\newblock \emph{IEEE Transactions on Systems, Man, and Cybernetics, Part B (Cybernetics)}, 35\penalty0 (5):\penalty0 894--904, 2005.

\bibitem[Chen et~al.(2024)Chen, Li, Wang, Xue, and Pang]{chen2024gennbv}
Xiao Chen, Quanyi Li, Tai Wang, Tianfan Xue, and Jiangmiao Pang.
\newblock Gennbv: Generalizable next-best-view policy for active 3d reconstruction.
\newblock In \emph{Proceedings of the IEEE/CVF Conference on Computer Vision and Pattern Recognition}, pages 16436--16445, 2024.

\bibitem[Connolly(1985)]{connolly1985determination}
Cl Connolly.
\newblock The determination of next best views.
\newblock In \emph{Proceedings. 1985 IEEE international conference on robotics and automation}, pages 432--435. IEEE, 1985.

\bibitem[Downs et~al.(2022)Downs, Francis, Koenig, Kinman, Hickman, Reymann, McHugh, and Vanhoucke]{downs2022google}
Laura Downs, Anthony Francis, Nate Koenig, Brandon Kinman, Ryan Hickman, Krista Reymann, Thomas~B McHugh, and Vincent Vanhoucke.
\newblock Google scanned objects: A high-quality dataset of 3d scanned household items.
\newblock In \emph{2022 International Conference on Robotics and Automation (ICRA)}, pages 2553--2560. IEEE, 2022.

\bibitem[Galceran and Carreras(2013)]{galceran2013survey}
Enric Galceran and Marc Carreras.
\newblock A survey on coverage path planning for robotics.
\newblock \emph{Robotics and Autonomous systems}, 61\penalty0 (12):\penalty0 1258--1276, 2013.

\bibitem[Ge et~al.(2023)Ge, Nah, Liu, Poon, Tao, Catanzaro, Jacobs, Huang, Liu, and Balaji]{ge2023preserve}
Songwei Ge, Seungjun Nah, Guilin Liu, Tyler Poon, Andrew Tao, Bryan Catanzaro, David Jacobs, Jia-Bin Huang, Ming-Yu Liu, and Yogesh Balaji.
\newblock Preserve your own correlation: A noise prior for video diffusion models.
\newblock In \emph{Proceedings of the IEEE/CVF International Conference on Computer Vision}, pages 22930--22941, 2023.

\bibitem[Girdhar et~al.(2023)Girdhar, Singh, Brown, Duval, Azadi, Rambhatla, Shah, Yin, Parikh, and Misra]{girdhar2023emu}
Rohit Girdhar, Mannat Singh, Andrew Brown, Quentin Duval, Samaneh Azadi, Sai~Saketh Rambhatla, Akbar Shah, Xi Yin, Devi Parikh, and Ishan Misra.
\newblock Emu video: Factorizing text-to-video generation by explicit image conditioning.
\newblock \emph{arXiv preprint arXiv:2311.10709}, 2023.

\bibitem[Guo et~al.(2023)Guo, Yang, Rao, Liang, Wang, Qiao, Agrawala, Lin, and Dai]{guo2023animatediff}
Yuwei Guo, Ceyuan Yang, Anyi Rao, Zhengyang Liang, Yaohui Wang, Yu Qiao, Maneesh Agrawala, Dahua Lin, and Bo Dai.
\newblock Animatediff: Animate your personalized text-to-image diffusion models without specific tuning.
\newblock \emph{arXiv preprint arXiv:2307.04725}, 2023.

\bibitem[He et~al.(2024)He, Xu, Guo, Wetzstein, Dai, Li, and Yang]{he2024cameractrl}
Hao He, Yinghao Xu, Yuwei Guo, Gordon Wetzstein, Bo Dai, Hongsheng Li, and Ceyuan Yang.
\newblock Cameractrl: Enabling camera control for text-to-video generation.
\newblock \emph{arXiv preprint arXiv:2404.02101}, 2024.

\bibitem[He et~al.(2016)He, Zhang, Ren, and Sun]{he2016deep}
Kaiming He, Xiangyu Zhang, Shaoqing Ren, and Jian Sun.
\newblock Deep residual learning for image recognition.
\newblock In \emph{Proceedings of the IEEE conference on computer vision and pattern recognition}, pages 770--778, 2016.

\bibitem[Ho et~al.(2022{\natexlab{a}})Ho, Chan, Saharia, Whang, Gao, Gritsenko, Kingma, Poole, Norouzi, Fleet, et~al.]{ho2022imagen}
Jonathan Ho, William Chan, Chitwan Saharia, Jay Whang, Ruiqi Gao, Alexey Gritsenko, Diederik~P Kingma, Ben Poole, Mohammad Norouzi, David~J Fleet, et~al.
\newblock Imagen video: High definition video generation with diffusion models.
\newblock \emph{arXiv preprint arXiv:2210.02303}, 2022{\natexlab{a}}.

\bibitem[Ho et~al.(2022{\natexlab{b}})Ho, Salimans, Gritsenko, Chan, Norouzi, and Fleet]{ho2022video}
Jonathan Ho, Tim Salimans, Alexey Gritsenko, William Chan, Mohammad Norouzi, and David~J Fleet.
\newblock Video diffusion models.
\newblock \emph{Advances in Neural Information Processing Systems}, 35:\penalty0 8633--8646, 2022{\natexlab{b}}.

\bibitem[Hong et~al.(2023)Hong, Seo, Shin, Hong, and Kim]{hong2023large}
Susung Hong, Junyoung Seo, Heeseong Shin, Sunghwan Hong, and Seungryong Kim.
\newblock Large language models are frame-level directors for zero-shot text-to-video generation.
\newblock In \emph{First Workshop on Controllable Video Generation@ ICML24}, 2023.

\bibitem[Huang et~al.(2024{\natexlab{a}})Huang, Yu, Chen, Geiger, and Gao]{huang20242d}
Binbin Huang, Zehao Yu, Anpei Chen, Andreas Geiger, and Shenghua Gao.
\newblock 2d gaussian splatting for geometrically accurate radiance fields.
\newblock In \emph{ACM SIGGRAPH 2024 Conference Papers}, pages 1--11, 2024{\natexlab{a}}.

\bibitem[Huang et~al.(2024{\natexlab{b}})Huang, Feng, Shi, Xu, Yu, and Yang]{huang2024free}
Hanzhuo Huang, Yufan Feng, Cheng Shi, Lan Xu, Jingyi Yu, and Sibei Yang.
\newblock Free-bloom: Zero-shot text-to-video generator with llm director and ldm animator.
\newblock \emph{Advances in Neural Information Processing Systems}, 36, 2024{\natexlab{b}}.

\bibitem[Jin et~al.(2023)Jin, Chen, R{\"u}ckin, and Popovi{\'c}]{jin2023neu}
Liren Jin, Xieyuanli Chen, Julius R{\"u}ckin, and Marija Popovi{\'c}.
\newblock Neu-nbv: Next best view planning using uncertainty estimation in image-based neural rendering.
\newblock In \emph{2023 IEEE/RSJ International Conference on Intelligent Robots and Systems (IROS)}, pages 11305--11312. IEEE, 2023.

\bibitem[Karur et~al.(2021)Karur, Sharma, Dharmatti, and Siegel]{karur2021survey}
Karthik Karur, Nitin Sharma, Chinmay Dharmatti, and Joshua~E Siegel.
\newblock A survey of path planning algorithms for mobile robots.
\newblock \emph{Vehicles}, 3\penalty0 (3):\penalty0 448--468, 2021.

\bibitem[Kerbl et~al.(2023)Kerbl, Kopanas, Leimk{\"u}hler, and Drettakis]{kerbl20233d}
Bernhard Kerbl, Georgios Kopanas, Thomas Leimk{\"u}hler, and George Drettakis.
\newblock {3D Gaussian} splatting for real-time radiance field rendering.
\newblock In \emph{SIGGRAPH}, pages 1--14, 2023.

\bibitem[Kriegel et~al.(2011)Kriegel, Bodenm{\"u}ller, Suppa, and Hirzinger]{kriegel2011surface}
Simon Kriegel, Tim Bodenm{\"u}ller, Michael Suppa, and Gerd Hirzinger.
\newblock A surface-based next-best-view approach for automated 3d model completion of unknown objects.
\newblock In \emph{2011 IEEE International Conference on Robotics and Automation}, pages 4869--4874. IEEE, 2011.

\bibitem[Li et~al.(2023{\natexlab{a}})Li, Tan, Zhang, Xu, Luan, Xu, Hong, Sunkavalli, Shakhnarovich, and Bi]{li2023instant3d}
Jiahao Li, Hao Tan, Kai Zhang, Zexiang Xu, Fujun Luan, Yinghao Xu, Yicong Hong, Kalyan Sunkavalli, Greg Shakhnarovich, and Sai Bi.
\newblock Instant3d: Fast text-to-3d with sparse-view generation and large reconstruction model.
\newblock \emph{arXiv preprint arXiv:2311.06214}, 2023{\natexlab{a}}.

\bibitem[Li et~al.(2024)Li, Liu, Chen, Liang, Chen, Tan, and Long]{li2024craftsman}
Weiyu Li, Jiarui Liu, Rui Chen, Yixun Liang, Xuelin Chen, Ping Tan, and Xiaoxiao Long.
\newblock Craftsman: High-fidelity mesh generation with 3d native generation and interactive geometry refiner.
\newblock \emph{arXiv preprint arXiv:2405.14979}, 2024.

\bibitem[Li et~al.(2023{\natexlab{b}})Li, Chu, Wu, Yuan, Liu, Zhang, Li, Feng, Ding, and Wang]{li2023videogen}
Xin Li, Wenqing Chu, Ye Wu, Weihang Yuan, Fanglong Liu, Qi Zhang, Fu Li, Haocheng Feng, Errui Ding, and Jingdong Wang.
\newblock Videogen: A reference-guided latent diffusion approach for high definition text-to-video generation.
\newblock \emph{arXiv preprint arXiv:2309.00398}, 2023{\natexlab{b}}.

\bibitem[Li et~al.(2023{\natexlab{c}})Li, M{\"u}ller, Evans, Taylor, Unberath, Liu, and Lin]{li2023neuralangelo}
Zhaoshuo Li, Thomas M{\"u}ller, Alex Evans, Russell~H Taylor, Mathias Unberath, Ming-Yu Liu, and Chen-Hsuan Lin.
\newblock Neuralangelo: High-fidelity neural surface reconstruction.
\newblock In \emph{Proceedings of the IEEE/CVF Conference on Computer Vision and Pattern Recognition}, pages 8456--8465, 2023{\natexlab{c}}.

\bibitem[Liu et~al.(2024{\natexlab{a}})Liu, Shi, Chen, Zhang, Xu, Wei, Chen, Zeng, Gu, and Su]{liu2024one}
Minghua Liu, Ruoxi Shi, Linghao Chen, Zhuoyang Zhang, Chao Xu, Xinyue Wei, Hansheng Chen, Chong Zeng, Jiayuan Gu, and Hao Su.
\newblock One-2-3-45++: Fast single image to 3d objects with consistent multi-view generation and 3d diffusion.
\newblock In \emph{Proceedings of the IEEE/CVF Conference on Computer Vision and Pattern Recognition}, pages 10072--10083, 2024{\natexlab{a}}.

\bibitem[Liu et~al.(2024{\natexlab{b}})Liu, Xu, Jin, Chen, Varma~T, Xu, and Su]{liu2024onev1}
Minghua Liu, Chao Xu, Haian Jin, Linghao Chen, Mukund Varma~T, Zexiang Xu, and Hao Su.
\newblock One-2-3-45: Any single image to 3d mesh in 45 seconds without per-shape optimization.
\newblock \emph{Advances in Neural Information Processing Systems}, 36, 2024{\natexlab{b}}.

\bibitem[Liu et~al.(2024{\natexlab{c}})Liu, Zeng, Wei, Shi, Chen, Xu, Zhang, Wang, Zhang, Liu, et~al.]{liu2024meshformer}
Minghua Liu, Chong Zeng, Xinyue Wei, Ruoxi Shi, Linghao Chen, Chao Xu, Mengqi Zhang, Zhaoning Wang, Xiaoshuai Zhang, Isabella Liu, et~al.
\newblock Meshformer: High-quality mesh generation with 3d-guided reconstruction model.
\newblock \emph{arXiv preprint arXiv:2408.10198}, 2024{\natexlab{c}}.

\bibitem[Liu et~al.(2023{\natexlab{a}})Liu, Wu, Van~Hoorick, Tokmakov, Zakharov, and Vondrick]{liu2023zero}
Ruoshi Liu, Rundi Wu, Basile Van~Hoorick, Pavel Tokmakov, Sergey Zakharov, and Carl Vondrick.
\newblock Zero-1-to-3: Zero-shot one image to 3d object.
\newblock In \emph{Proceedings of the IEEE/CVF international conference on computer vision}, pages 9298--9309, 2023{\natexlab{a}}.

\bibitem[Liu et~al.(2022)Liu, Lin, Hu, Xie, Fu, Zhang, and Huang]{liu2022learning}
Yilin Liu, Liqiang Lin, Yue Hu, Ke Xie, Chi-Wing Fu, Hao Zhang, and Hui Huang.
\newblock Learning reconstructability for drone aerial path planning.
\newblock \emph{ACM Transactions on Graphics (TOG)}, 41\penalty0 (6):\penalty0 1--17, 2022.

\bibitem[Liu et~al.(2023{\natexlab{b}})Liu, Lin, Zeng, Long, Liu, Komura, and Wang]{liu2023syncdreamer}
Yuan Liu, Cheng Lin, Zijiao Zeng, Xiaoxiao Long, Lingjie Liu, Taku Komura, and Wenping Wang.
\newblock Syncdreamer: Generating multiview-consistent images from a single-view image.
\newblock \emph{arXiv preprint arXiv:2309.03453}, 2023{\natexlab{b}}.

\bibitem[Long et~al.(2022)Long, Lin, Wang, Komura, and Wang]{long2022sparseneus}
Xiaoxiao Long, Cheng Lin, Peng Wang, Taku Komura, and Wenping Wang.
\newblock {SparseNeuS}: Fast generalizable neural surface reconstruction from sparse views.
\newblock In \emph{Computer Vision--ECCV 2022: 17th European Conference, Tel Aviv, Israel, October 23--27, 2022, Proceedings, Part XXXII}, pages 210--227. Springer, 2022.

\bibitem[Long et~al.(2024)Long, Guo, Lin, Liu, Dou, Liu, Ma, Zhang, Habermann, Theobalt, et~al.]{long2024wonder3d}
Xiaoxiao Long, Yuan-Chen Guo, Cheng Lin, Yuan Liu, Zhiyang Dou, Lingjie Liu, Yuexin Ma, Song-Hai Zhang, Marc Habermann, Christian Theobalt, et~al.
\newblock Wonder3d: Single image to 3d using cross-domain diffusion.
\newblock In \emph{Proceedings of the IEEE/CVF Conference on Computer Vision and Pattern Recognition}, pages 9970--9980, 2024.

\bibitem[Mahdavi-Amiri et~al.(2020)Mahdavi-Amiri, Yu, Zhao, Schulz, and Zhang]{VDACAli}
Ali Mahdavi-Amiri, Fenggen Yu, Haisen Zhao, Adriana Schulz, and Hao Zhang.
\newblock Vdac: volume decompose-and-carve for subtractive manufacturing.
\newblock \emph{ACM Trans. Graph.}, 39\penalty0 (6), 2020.

\bibitem[Maldonado et~al.(2016)Maldonado, Hadfield, Pugeault, and Bowden]{maldonado2016next}
Oscar~Mendez Maldonado, Simon Hadfield, Nicolas Pugeault, and Richard Bowden.
\newblock Next-best stereo: extending next best view optimisation for collaborative sensors.
\newblock \emph{Proceedings of BMVC 2016}, 2016.

\bibitem[Massios et~al.(1998)Massios, Fisher, et~al.]{massios1998best}
Nikolaos~A Massios, Robert~B Fisher, et~al.
\newblock \emph{A best next view selection algorithm incorporating a quality criterion}.
\newblock Citeseer, 1998.

\bibitem[Melas-Kyriazi et~al.(2024)Melas-Kyriazi, Laina, Rupprecht, Neverova, Vedaldi, Gafni, and Kokkinos]{melas20243d}
Luke Melas-Kyriazi, Iro Laina, Christian Rupprecht, Natalia Neverova, Andrea Vedaldi, Oran Gafni, and Filippos Kokkinos.
\newblock Im-3d: Iterative multiview diffusion and reconstruction for high-quality 3d generation.
\newblock \emph{arXiv preprint arXiv:2402.08682}, 2024.

\bibitem[Mildenhall et~al.(2020)Mildenhall, Srinivasan, Tancik, Barron, Ramamoorthi, and Ng]{nerf}
Ben Mildenhall, Pratul~P. Srinivasan, Matthew Tancik, Jonathan~T. Barron, Ravi Ramamoorthi, and Ren Ng.
\newblock {NeRF}: Representing scenes as neural radiance fields for view synthesis.
\newblock In \emph{ECCV}, 2020.

\bibitem[Oquab et~al.(2023)Oquab, Darcet, Moutakanni, Vo, Szafraniec, Khalidov, Fernandez, Haziza, Massa, El-Nouby, et~al.]{oquab2023dinov2}
Maxime Oquab, Timoth{\'e}e Darcet, Th{\'e}o Moutakanni, Huy Vo, Marc Szafraniec, Vasil Khalidov, Pierre Fernandez, Daniel Haziza, Francisco Massa, Alaaeldin El-Nouby, et~al.
\newblock Dinov2: Learning robust visual features without supervision.
\newblock \emph{arXiv preprint arXiv:2304.07193}, 2023.

\bibitem[Peralta et~al.(2020)Peralta, Casimiro, Nilles, Aguilar, Atienza, and Cajote]{peralta2020next}
Daryl Peralta, Joel Casimiro, Aldrin~Michael Nilles, Justine~Aletta Aguilar, Rowel Atienza, and Rhandley Cajote.
\newblock Next-best view policy for 3d reconstruction.
\newblock In \emph{Computer Vision--ECCV 2020 Workshops: Glasgow, UK, August 23--28, 2020, Proceedings, Part IV 16}, pages 558--573. Springer, 2020.

\bibitem[Pito(1999)]{pito1999solution}
Richard Pito.
\newblock A solution to the next best view problem for automated surface acquisition.
\newblock \emph{IEEE Transactions on pattern analysis and machine intelligence}, 21\penalty0 (10):\penalty0 1016--1030, 1999.

\bibitem[Qin et~al.(2020)Qin, Zhang, Huang, Dehghan, Zaiane, and Jagersand]{qin2020u2}
Xuebin Qin, Zichen Zhang, Chenyang Huang, Masood Dehghan, Osmar~R Zaiane, and Martin Jagersand.
\newblock U2-net: Going deeper with nested u-structure for salient object detection.
\newblock \emph{Pattern Recognition}, 106:\penalty0 107404, 2020.

\bibitem[Radford et~al.(2021)Radford, Kim, Hallacy, Ramesh, Goh, Agarwal, Sastry, Askell, Mishkin, Clark, et~al.]{radford2021learning}
Alec Radford, Jong~Wook Kim, Chris Hallacy, Aditya Ramesh, Gabriel Goh, Sandhini Agarwal, Girish Sastry, Amanda Askell, Pamela Mishkin, Jack Clark, et~al.
\newblock Learning transferable visual models from natural language supervision.
\newblock In \emph{International conference on machine learning}, pages 8748--8763. PmLR, 2021.

\bibitem[Roberts and Hanrahan(2016)]{roberts2016generating}
Mike Roberts and Pat Hanrahan.
\newblock Generating dynamically feasible trajectories for quadrotor cameras.
\newblock \emph{ACM Transactions on Graphics (TOG)}, 35\penalty0 (4):\penalty0 1--11, 2016.

\bibitem[Roberts et~al.(2017)Roberts, Dey, Truong, Sinha, Shah, Kapoor, Hanrahan, and Joshi]{roberts2017submodular}
Mike Roberts, Debadeepta Dey, Anh Truong, Sudipta Sinha, Shital Shah, Ashish Kapoor, Pat Hanrahan, and Neel Joshi.
\newblock Submodular trajectory optimization for aerial 3d scanning.
\newblock In \emph{Proceedings of the IEEE International Conference on Computer Vision}, pages 5324--5333, 2017.

\bibitem[Rombach et~al.(2022{\natexlab{a}})Rombach, Blattmann, Lorenz, Esser, and Ommer]{rombach2022high}
Robin Rombach, Andreas Blattmann, Dominik Lorenz, Patrick Esser, and Bj{\"o}rn Ommer.
\newblock High-resolution image synthesis with latent diffusion models.
\newblock In \emph{Proceedings of the IEEE/CVF conference on computer vision and pattern recognition}, pages 10684--10695, 2022{\natexlab{a}}.

\bibitem[Rombach et~al.(2022{\natexlab{b}})Rombach, Blattmann, Lorenz, Esser, and Ommer]{rombach2022sd}
Robin Rombach, Andreas Blattmann, Dominik Lorenz, Patrick Esser, and Bjorn Ommer.
\newblock High-resolution image synthesis with latent diffusion models.
\newblock In \emph{CVPR}, pages 10684--10695, 2022{\natexlab{b}}.

\bibitem[Shi et~al.(2023)Shi, Chen, Zhang, Liu, Xu, Wei, Chen, Zeng, and Su]{shi2023zero123++}
Ruoxi Shi, Hansheng Chen, Zhuoyang Zhang, Minghua Liu, Chao Xu, Xinyue Wei, Linghao Chen, Chong Zeng, and Hao Su.
\newblock Zero123++: a single image to consistent multi-view diffusion base model.
\newblock \emph{arXiv preprint arXiv:2310.15110}, 2023.

\bibitem[Simonyan and Zisserman(2014)]{simonyan2014very}
Karen Simonyan and Andrew Zisserman.
\newblock Very deep convolutional networks for large-scale image recognition.
\newblock \emph{arXiv preprint arXiv:1409.1556}, 2014.

\bibitem[Singer et~al.(2022)Singer, Polyak, Hayes, Yin, An, Zhang, Hu, Yang, Ashual, Gafni, et~al.]{singer2022make}
Uriel Singer, Adam Polyak, Thomas Hayes, Xi Yin, Jie An, Songyang Zhang, Qiyuan Hu, Harry Yang, Oron Ashual, Oran Gafni, et~al.
\newblock Make-a-video: Text-to-video generation without text-video data.
\newblock \emph{arXiv preprint arXiv:2209.14792}, 2022.

\bibitem[Tang et~al.(2024{\natexlab{a}})Tang, Chen, Chen, Wang, Zeng, and Liu]{tang2025lgm}
Jiaxiang Tang, Zhaoxi Chen, Xiaokang Chen, Tengfei Wang, Gang Zeng, and Ziwei Liu.
\newblock {LGM}: Large multi-view gaussian model for high-resolution 3d content creation.
\newblock In \emph{European Conference on Computer Vision}, pages 1--18. Springer, 2024{\natexlab{a}}.

\bibitem[Tang et~al.(2023)Tang, Zhang, Chen, Wang, and Furukawa]{tang2023mvdiffusionenablingholisticmultiview}
Shitao Tang, Fuyang Zhang, Jiacheng Chen, Peng Wang, and Yasutaka Furukawa.
\newblock Mvdiffusion: Enabling holistic multi-view image generation with correspondence-aware diffusion, 2023.

\bibitem[Tang et~al.(2024{\natexlab{b}})Tang, Chen, Wang, Tang, Zhang, Fan, Chandra, Furukawa, and Ranjan]{tang2024mvdiffusion++}
Shitao Tang, Jiacheng Chen, Dilin Wang, Chengzhou Tang, Fuyang Zhang, Yuchen Fan, Vikas Chandra, Yasutaka Furukawa, and Rakesh Ranjan.
\newblock Mvdiffusion++: A dense high-resolution multi-view diffusion model for single or sparse-view 3d object reconstruction.
\newblock \emph{arXiv preprint arXiv:2402.12712}, 2024{\natexlab{b}}.

\bibitem[Tatarchenko et~al.(2019)Tatarchenko, Richter, Ranftl, Li, Koltun, and Brox]{tatarchenko2019single}
Maxim Tatarchenko, Stephan~R Richter, Ren{\'e} Ranftl, Zhuwen Li, Vladlen Koltun, and Thomas Brox.
\newblock What do single-view 3d reconstruction networks learn?
\newblock In \emph{CVPR}, pages 3405--3414, 2019.

\bibitem[Team(2025)]{hunyuan3d22025tencent}
Tencent~Hunyuan3D Team.
\newblock Hunyuan3d 2.0: Scaling diffusion models for high resolution textured 3d assets generation, 2025.

\bibitem[Vasquez-Gomez et~al.(2014)Vasquez-Gomez, Sucar, Murrieta-Cid, and Lopez-Damian]{vasquez2014volumetric}
J~Irving Vasquez-Gomez, L~Enrique Sucar, Rafael Murrieta-Cid, and Efrain Lopez-Damian.
\newblock Volumetric next-best-view planning for 3d object reconstruction with positioning error.
\newblock \emph{International Journal of Advanced Robotic Systems}, 11\penalty0 (10):\penalty0 159, 2014.

\bibitem[Voleti et~al.(2024)Voleti, Yao, Boss, Letts, Pankratz, Tochilkin, Laforte, Rombach, and Jampani]{voleti2024sv3d}
Vikram Voleti, Chun-Han Yao, Mark Boss, Adam Letts, David Pankratz, Dmitry Tochilkin, Christian Laforte, Robin Rombach, and Varun Jampani.
\newblock Sv3d: Novel multi-view synthesis and 3d generation from a single image using latent video diffusion.
\newblock \emph{arXiv preprint arXiv:2403.12008}, 2024.

\bibitem[Vora et~al.(2023)Vora, Patil, and Zhang]{vora2023DiViNet}
Aditya Vora, Akshay~Gadi Patil, and Hao Zhang.
\newblock {DiViNet}: Artistic typography via discriminated and stylized diffusion.
\newblock In \emph{Proc. of NeurIPS}, 2023.

\bibitem[Wang et~al.(2021)Wang, Liu, Liu, Theobalt, Komura, and Wang]{wang2021neus}
Peng Wang, Lingjie Liu, Yuan Liu, Christian Theobalt, Taku Komura, and Wenping Wang.
\newblock {NeUS}: Learning neural implicit surfaces by volume rendering for multi-view reconstruction.
\newblock \emph{arXiv preprint arXiv:2106.10689}, 2021.

\bibitem[Wang et~al.(2024{\natexlab{a}})Wang, Bao, Weng, Feng, Yin, Yang, Zhang, Dai, Zhao, Wang, et~al.]{wang2024microcinema}
Yanhui Wang, Jianmin Bao, Wenming Weng, Ruoyu Feng, Dacheng Yin, Tao Yang, Jingxu Zhang, Qi Dai, Zhiyuan Zhao, Chunyu Wang, et~al.
\newblock Microcinema: A divide-and-conquer approach for text-to-video generation.
\newblock In \emph{Proceedings of the IEEE/CVF Conference on Computer Vision and Pattern Recognition}, pages 8414--8424, 2024{\natexlab{a}}.

\bibitem[Wang et~al.(2024{\natexlab{b}})Wang, Lira, Wang, Mahdavi-Amiri, and Zhang]{wang2024slice3d}
Yizhi Wang, Wallace Lira, Wenqi Wang, Ali Mahdavi-Amiri, and Hao Zhang.
\newblock Slice3d: Multi-slice occlusion-revealing single view 3d reconstruction.
\newblock In \emph{Proceedings of the IEEE/CVF Conference on Computer Vision and Pattern Recognition}, pages 9881--9891, 2024{\natexlab{b}}.

\bibitem[Wang et~al.(2004)Wang, Bovik, Sheikh, and Simoncelli]{wang2004image}
Zhou Wang, Alan~C Bovik, Hamid~R Sheikh, and Eero~P Simoncelli.
\newblock Image quality assessment: from error visibility to structural similarity.
\newblock \emph{IEEE transactions on image processing}, 13\penalty0 (4):\penalty0 600--612, 2004.

\bibitem[Wang et~al.(2024{\natexlab{c}})Wang, Wang, Chen, Xiang, Chen, Yu, Li, Su, and Zhu]{wang2024crm}
Zhengyi Wang, Yikai Wang, Yifei Chen, Chendong Xiang, Shuo Chen, Dajiang Yu, Chongxuan Li, Hang Su, and Jun Zhu.
\newblock Crm: Single image to 3d textured mesh with convolutional reconstruction model.
\newblock \emph{arXiv preprint arXiv:2403.05034}, 2024{\natexlab{c}}.

\bibitem[Watson et~al.(2023)Watson, Chan, Martin-Brualla, Ho, Tagliasacchi, and Norouzi]{watson2023novel}
Daniel Watson, William Chan, Ricardo Martin-Brualla, Jonathan Ho, Andrea Tagliasacchi, and Mohammad Norouzi.
\newblock Novel view synthesis with diffusion models.
\newblock In \emph{ICLR}, 2023.

\bibitem[Wu et~al.(2023)Wu, Ge, Wang, Lei, Gu, Shi, Hsu, Shan, Qie, and Shou]{wu2023tune}
Jay~Zhangjie Wu, Yixiao Ge, Xintao Wang, Stan~Weixian Lei, Yuchao Gu, Yufei Shi, Wynne Hsu, Ying Shan, Xiaohu Qie, and Mike~Zheng Shou.
\newblock Tune-a-video: One-shot tuning of image diffusion models for text-to-video generation.
\newblock In \emph{Proceedings of the IEEE/CVF International Conference on Computer Vision}, pages 7623--7633, 2023.

\bibitem[Wu et~al.(2014)Wu, Sun, Long, Huang, Cohen-Or, Gong, Deussen, and Chen]{wu2014autoscanning}
Shihao Wu, Wei Sun, Pinxin Long, Hui Huang, Daniel Cohen-Or, Minglun Gong, Oliver Deussen, and Baoquan Chen.
\newblock Quality-driven poisson-guided autoscanning.
\newblock \emph{ACM Transactions on Graphics (Proc. SIGGRAPH Asia)}, 33:\penalty0 203:1–203:12, 2014.

\bibitem[Wu et~al.(2024)Wu, Lin, Zhang, Zeng, Xu, Torr, Cao, and Yao]{wu2024direct3d}
Shuang Wu, Youtian Lin, Feihu Zhang, Yifei Zeng, Jingxi Xu, Philip Torr, Xun Cao, and Yao Yao.
\newblock Direct3d: Scalable image-to-3d generation via 3d latent diffusion transformer.
\newblock \emph{arXiv preprint arXiv:2405.14832}, 2024.

\bibitem[Xiang et~al.(2025)Xiang, Lv, Xu, Deng, Wang, Zhang, Chen, Tong, and Yang]{xiang2024structured}
Jianfeng Xiang, Zelong Lv, Sicheng Xu, Yu Deng, Ruicheng Wang, Bowen Zhang, Dong Chen, Xin Tong, and Jiaolong Yang.
\newblock Structured 3d latents for scalable and versatile 3d generation.
\newblock In \emph{CVPR}, 2025.

\bibitem[Xu et~al.(2024{\natexlab{a}})Xu, Nie, Liu, Liu, Kautz, Wang, and Vahdat]{xu2024camco}
Dejia Xu, Weili Nie, Chao Liu, Sifei Liu, Jan Kautz, Zhangyang Wang, and Arash Vahdat.
\newblock Camco: Camera-controllable 3d-consistent image-to-video generation.
\newblock \emph{arXiv preprint arXiv:2406.02509}, 2024{\natexlab{a}}.

\bibitem[Xu et~al.(2024{\natexlab{b}})Xu, Cheng, Gao, Wang, Gao, and Shan]{xu2024instantmesh}
Jiale Xu, Weihao Cheng, Yiming Gao, Xintao Wang, Shenghua Gao, and Ying Shan.
\newblock Instantmesh: Efficient 3d mesh generation from a single image with sparse-view large reconstruction models.
\newblock \emph{arXiv preprint arXiv:2404.07191}, 2024{\natexlab{b}}.

\bibitem[Yang et~al.(2024)Yang, Zuo, Ramasinghe, Bazzani, Avraham, and van~den Hengel]{yang2024viewfusion}
Xianghui Yang, Yan Zuo, Sameera Ramasinghe, Loris Bazzani, Gil Avraham, and Anton van~den Hengel.
\newblock Viewfusion: Towards multi-view consistency via interpolated denoising.
\newblock In \emph{Proceedings of the IEEE/CVF Conference on Computer Vision and Pattern Recognition}, pages 9870--9880, 2024.

\bibitem[Yu et~al.(2024)Yu, Xing, Yuan, Hu, Li, Huang, Gao, Wong, Shan, and Tian]{yu2024viewcrafter}
Wangbo Yu, Jinbo Xing, Li Yuan, Wenbo Hu, Xiaoyu Li, Zhipeng Huang, Xiangjun Gao, Tien-Tsin Wong, Ying Shan, and Yonghong Tian.
\newblock Viewcrafter: Taming video diffusion models for high-fidelity novel view synthesis.
\newblock \emph{arXiv preprint arXiv:2409.02048}, 2024.

\bibitem[Zeng et~al.(2024)Zeng, Wei, Zheng, Zou, Wei, Zhang, and Li]{zeng2024make}
Yan Zeng, Guoqiang Wei, Jiani Zheng, Jiaxin Zou, Yang Wei, Yuchen Zhang, and Hang Li.
\newblock Make pixels dance: High-dynamic video generation.
\newblock In \emph{Proceedings of the IEEE/CVF Conference on Computer Vision and Pattern Recognition}, pages 8850--8860, 2024.

\bibitem[Zhang et~al.(2024)Zhang, Wang, Zhang, Qiu, Pang, Jiang, Yang, Xu, and Yu]{zhang2024clay}
Longwen Zhang, Ziyu Wang, Qixuan Zhang, Qiwei Qiu, Anqi Pang, Haoran Jiang, Wei Yang, Lan Xu, and Jingyi Yu.
\newblock Clay: A controllable large-scale generative model for creating high-quality 3d assets.
\newblock \emph{ACM Transactions on Graphics (TOG)}, 43\penalty0 (4):\penalty0 1--20, 2024.

\bibitem[Zhang et~al.(2018)Zhang, Isola, Efros, Shechtman, and Wang]{zhang2018unreasonable}
Richard Zhang, Phillip Isola, Alexei~A Efros, Eli Shechtman, and Oliver Wang.
\newblock The unreasonable effectiveness of deep features as a perceptual metric.
\newblock In \emph{Proceedings of the IEEE conference on computer vision and pattern recognition}, pages 586--595, 2018.

\bibitem[Zhang et~al.(2023)Zhang, Wang, Zhang, Zhao, Yuan, Qin, Wang, Zhao, and Zhou]{zhang2023i2vgen}
Shiwei Zhang, Jiayu Wang, Yingya Zhang, Kang Zhao, Hangjie Yuan, Zhiwu Qin, Xiang Wang, Deli Zhao, and Jingren Zhou.
\newblock I2vgen-xl: High-quality image-to-video synthesis via cascaded diffusion models.
\newblock \emph{arXiv preprint arXiv:2311.04145}, 2023.

\bibitem[Zhao et~al.(2023)Zhao, Liu, Chen, Zeng, Wang, Cheng, Fu, Chen, Yu, and Gao]{zhao2023michelangelo}
Zibo Zhao, Wen Liu, Xin Chen, Xianfang Zeng, Rui Wang, Pei Cheng, Bin Fu, Tao Chen, Gang Yu, and Shenghua Gao.
\newblock Michelangelo: Conditional 3d shape generation based on shape-image-text aligned latent representation.
\newblock \emph{Advances in neural information processing systems}, 36:\penalty0 73969--73982, 2023.

\bibitem[Zhou et~al.(2022)Zhou, Wang, Yan, Lv, Zhu, and Feng]{zhou2022magicvideo}
Daquan Zhou, Weimin Wang, Hanshu Yan, Weiwei Lv, Yizhe Zhu, and Jiashi Feng.
\newblock Magicvideo: Efficient video generation with latent diffusion models.
\newblock \emph{arXiv preprint arXiv:2211.11018}, 2022.

\end{thebibliography}
}

\clearpage
\setcounter{page}{1}
\maketitlesupplementary


\section{Additional Results}
\begin{figure*}[!t]
    \centering
    \includegraphics[width=\textwidth]{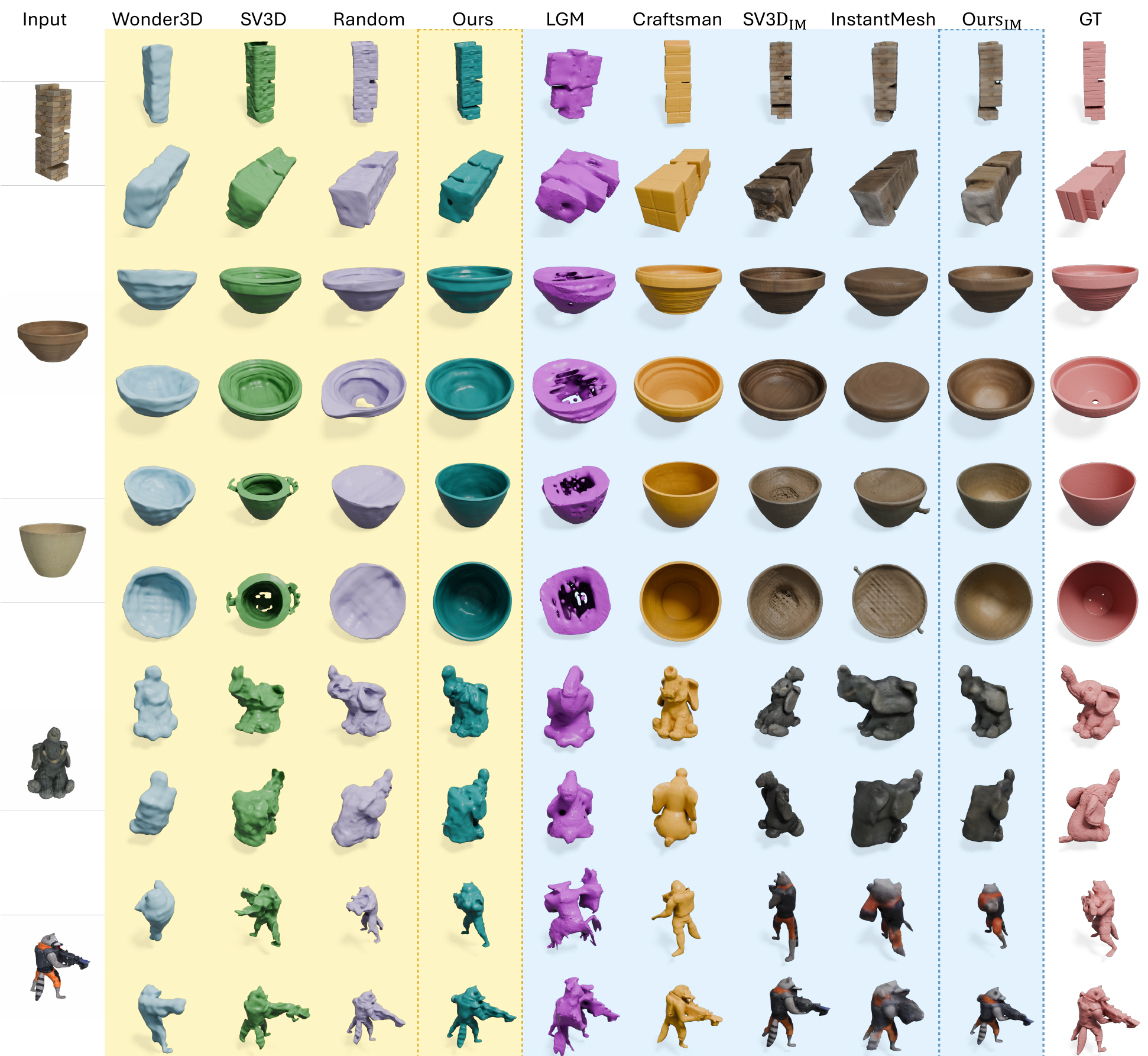}
\caption{Additional qualitative visual comparisons between single-view 3D reconstruction methods on the GSO dataset. Please zoom in for a closer inspection. Meshes reconstructed with NeUS and LRM are in yellow and blue blocks, respectively. Our slice-guided approach demonstrates superior reconstruction of occluded regions, including bowl bottoms and extended features like elephant trunk. Please zoom in for a detailed comparison of the reconstruction quality.}
\vspace{-0.2cm}
\label{fig:additional_result00}
\end{figure*}

\begin{figure*}[!t]
    \centering
    \includegraphics[width=\textwidth]{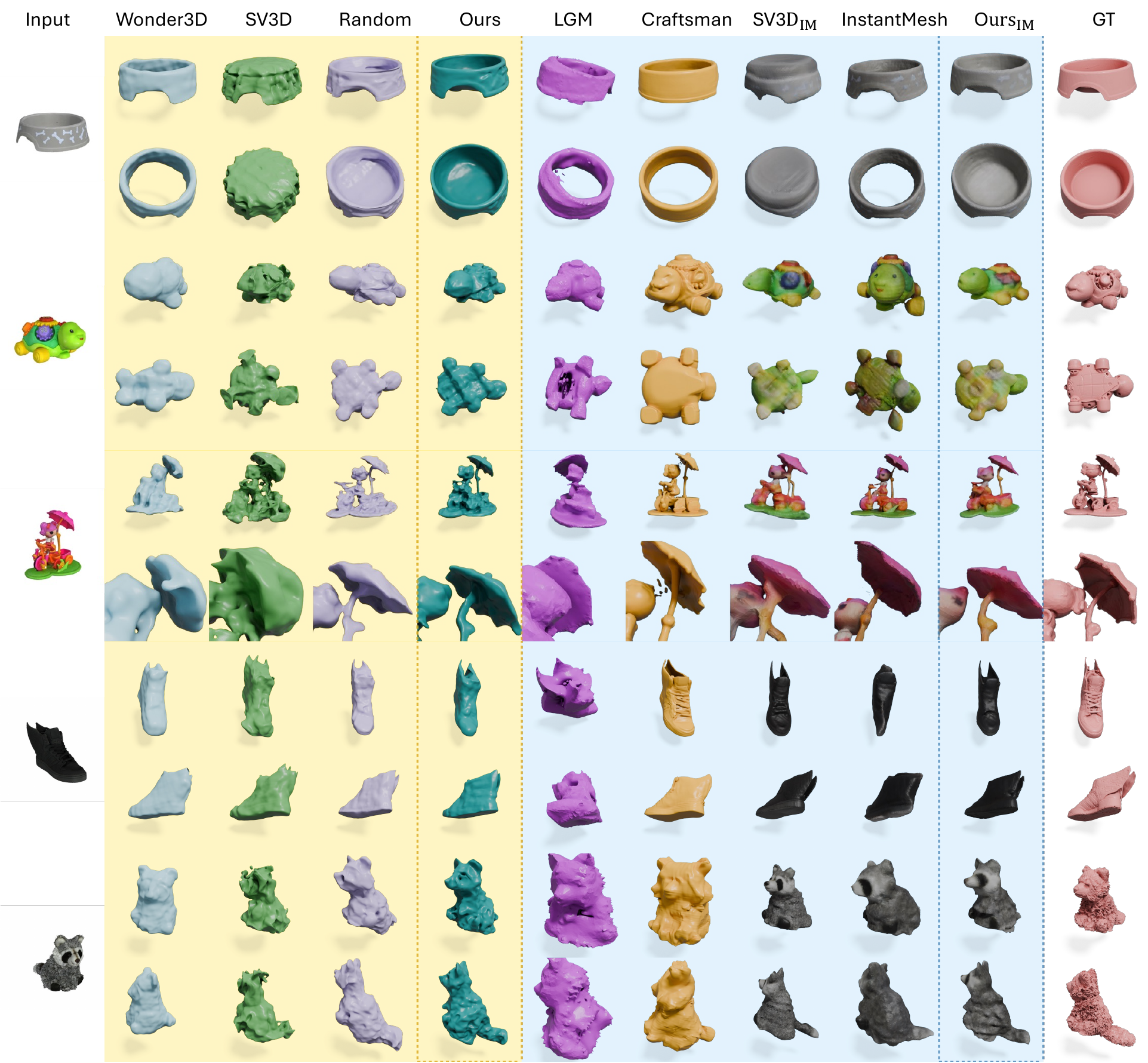}
\caption{Additional qualitative visual comparisons between single-view 3D reconstruction methods on the GSO dataset. Please zoom in for a closer inspection. Meshes reconstructed with NeUS and LRM are in yellow and blue blocks, respectively. Our slice-guided approach demonstrates superior reconstruction of occluded regions, including bowl bottoms, hidden wings and raccoon tails. Please zoom in for a detailed comparison of the reconstruction quality.}
\vspace{-0.2cm}
\label{fig:additional_result01}
\end{figure*}

We provide additional qualitative results from the GSO dataset in \Cref{fig:additional_result00,fig:additional_result01}.

\section{Estimation of Camera Poses and Object Bounding Boxes}
%

We train a network to predict the camera elevation and distance from a single image using Objaverse data. Specifically, we render different views of an object and obtain the ground-truth (GT) camera parameters for each rendered view. The network backbone is a ResNet50~\cite{he2016deep}, with fully connected layers adapted to predict (1) the elevation $e_0$ and the camera distance $d$; and (2) the edge lengths in three dimensions $(w, h, l)$ of the bounding box.

Instead of directly aligning the predicted elevation $e'_0$ and distance $d'$ with their GT counterparts $e_0$ and $d$, we adopt an alternative approach. From each shape, we sample a point cloud $P$ with $1,024$ points. The camera elevation and distance are transformed into a matrix $\Theta$, which performs the coordinate transformation from world coordinates to camera coordinates.
For the predicted and ground-truth transformation matrices $\Theta'$ and $\Theta$, the objective function minimizes the discrepancy between the transformed point clouds:

\begin{equation}
    L_{cam} = ||P\Theta' - P\Theta||_{2},
\end{equation}
which measures the error by comparing the rotated point clouds.

For the bounding boxes prediction, we directly use MSE:

\begin{equation}
    \mathcal{L}_{bbox} = || w - w'||_{2} + || l - l'||_{2} + || h - h'||_{2},
\end{equation}
where $(w', h', l')$ are the predicted edge lengths.

The overall training loss $\mathcal{L}$ combines the bounding box loss $\mathcal{L}_{bbox}$ and the camera alignment loss $L_{cam}$:

\begin{equation}
\mathcal{L} = \mathcal{L}_{bbox} + L_{cam}.
\end{equation}

\section{Robustness Against Camera Pose Errors}

\begin{figure}[!t]
    \centering
    \includegraphics[width=0.99\columnwidth]{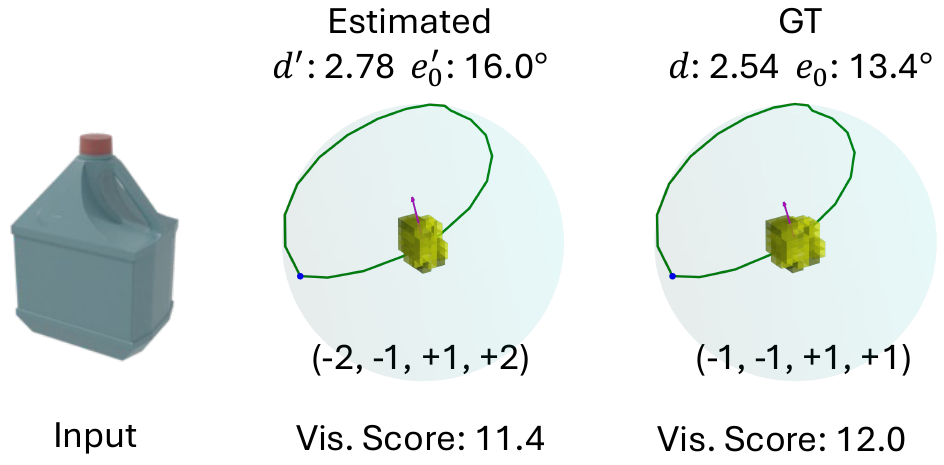}
\caption{
Robustness to camera pose prediction errors. The left and right show our predicted camera trajectories using predicted and GT camera poses, respectively. The numbers in brackets show the elevation changes in each segment of the orbit. "Vis. Score" denotes the visibility score defined in Equation (1) from the main paper.}
\label{fig:cmr_est_robst}
\end{figure}

Our method demonstrates notable robustness to errors in camera pose estimation. For example, as shown in Fig.~\ref{fig:cmr_est_robst}, the predicted camera distance and elevation deviate somewhat from the GT values. Despite this, our approach produces a trajectory closely resembling the one derived from GT camera poses, with only a minor reduction in the visibility score defined in Equation 2 of the main paper.

\section{Visibility Set Calculation}
To calculate the visibility set $\psi(\pi(t))$ in Equation (1) from the main paper, we perform two key checks for each difference block: (1) a field of view check and (2) an occlusion check. The field of view check ensures blocks lie within the camera's viewing angle (33.6°), computed as the angle between the camera direction and the vector from camera to block center. The occlusion check determines if any other block intersects the line segment from the camera to the target block's center using a ray-box intersection algorithm. A block is considered visible only if it passes both checks. This visibility function allows us to evaluate each candidate trajectory by summing the weights of all visible blocks across all camera positions, leading to the selection of the optimal trajectory $\pi^*$ that maximizes the visibility of important difference regions.

\section{Calculation of Random Trajectory}
Following the same setup as our adaptive trajectories, we apply a uniform azimuth angle change of $18^\circ$ and divide the camera trajectory orbit into four segments. Instead of setting a constant elevation increment stepsize within each segment, we randomly sample an incrementation step with $[-5^\circ, 5^\circ]$ for each time of elevation change. To ensure the final trajectory forms a close loop, we apply the same mirroring and negating mechanism between segments.

\section{Semantic Difference Map}
\begin{figure}
    \centering
    \includegraphics[width=\linewidth]{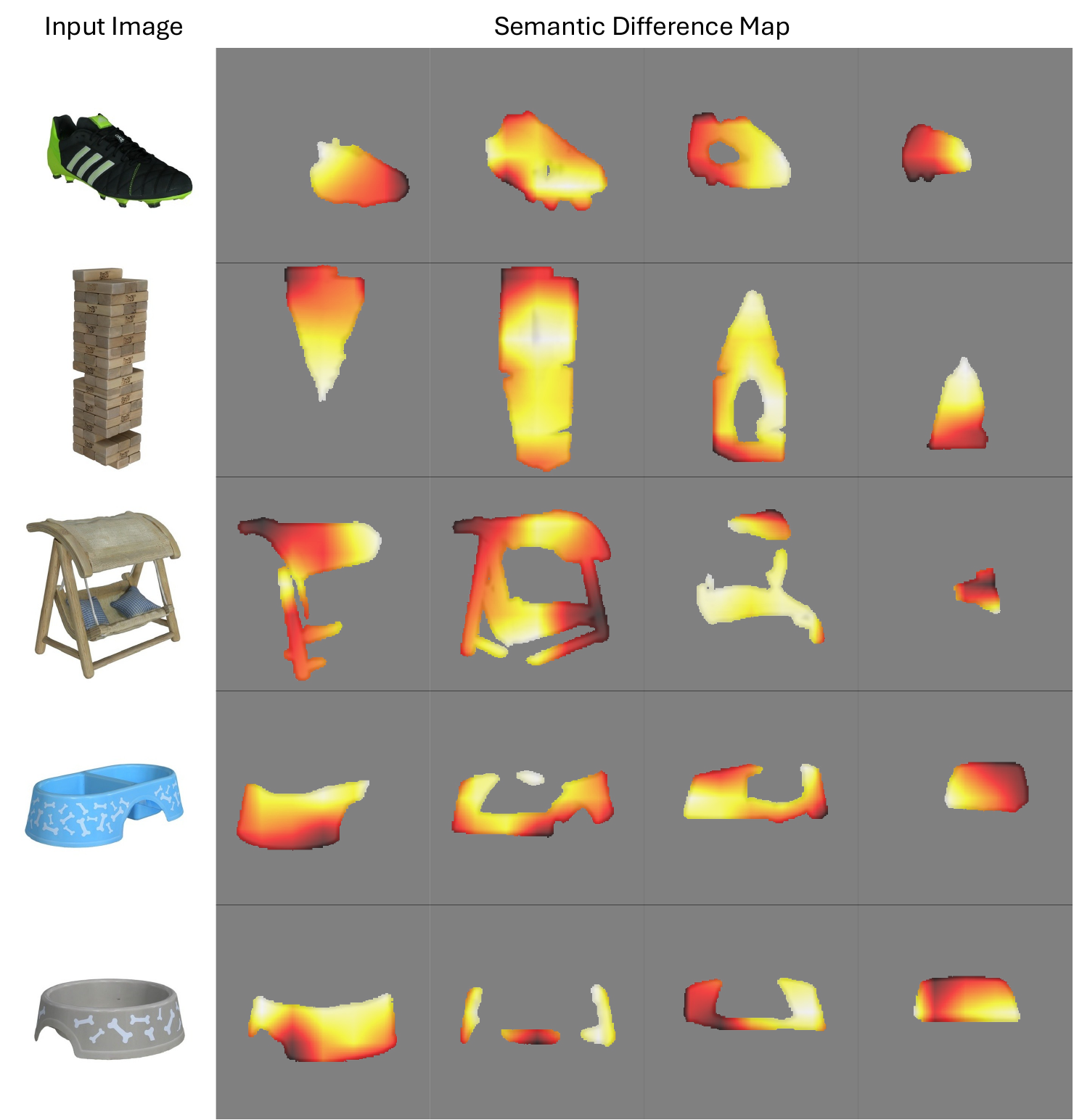}
\vspace{-1em}
\caption{Visualization of semantic difference map computed form feature comparison between input image and the generated slice images by Slice3D~\cite{wang2024slice3d}. The brighter color indicates higher semantic difference, and the darker colour indicates lower semantic differences.
}
\label{fig:vis_of_semantic_difference_map}
\end{figure}

We show additional examples on semantic difference map obatained by comparing feature differences between input image and the generated slices from Slice3D~\cite{wang2024slice3d}, as shown in \Cref{fig:vis_of_semantic_difference_map}.

\section{Alternative Feature Encoders}

In our primary experiments, we utilized VGG16~\cite{simonyan2014very} as the feature extractor to quantify semantic differences between slice images and the input image. To evaluate the impact of feature encoder architecture, we conducted ablation studies with DINOv2~\cite{oquab2023dinov2} and CLIP~\cite{radford2021learning} as alternative feature extractors. For the DINOv2 implementation, we extracted features from the final intermediate layer, resulting in a feature representation $\phi(\cdot)\in \mathbb{R}^{1536\times 16\times 16}$ while CLIP features remain in the same resolution as VGG. These features were then used to construct semantic difference maps following the same methodology applied to VGG features.

\begin{figure*}[!t]
    \centering
    \includegraphics[width=0.8\textwidth]{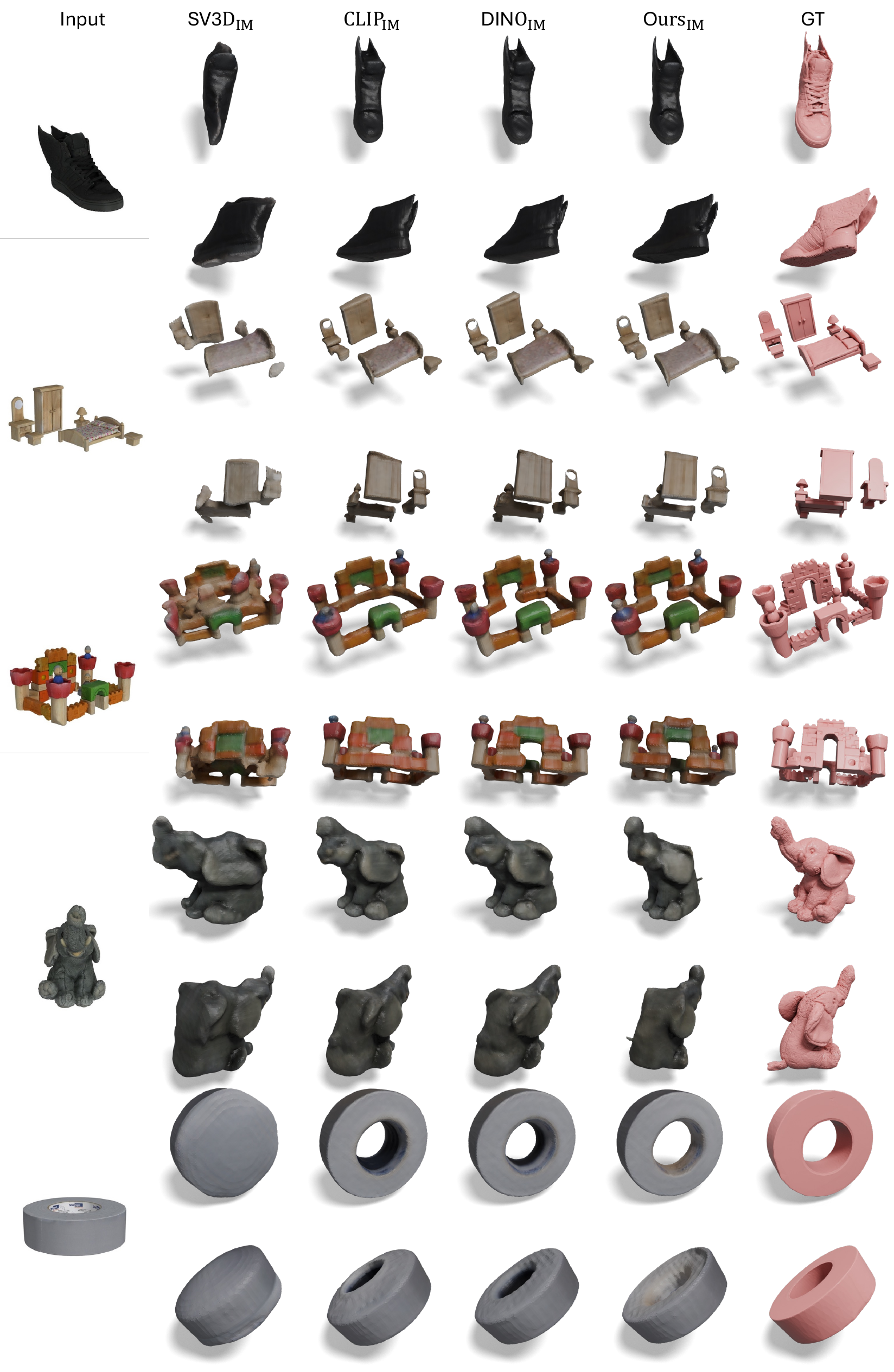}
\vspace{-0.5em}
\caption{Qualitative visual comparisons between 3D reconstruction with static camera trajectory (SV3D), adaptive camera trajectory with CLIP~\cite{radford2021learning}, DINOv2~\cite{oquab2023dinov2} (DINO) and VGG16 (Ours) as the feature extractor on the GSO dataset. Please zoom in for a closer inspection. All meshes are reconstructed with LRM. It can be seen that adaptive camera trajectory provides additional geometric awareness in comparison to static camera trajectory, regardless.}
\label{fig:dino_ablation}
\end{figure*}

Qualitative and quantitative comparisons between static camera trajectory (SV3D), adaptive trajectory with CLIP, DINOv2, and VGG are presented in \Cref{fig:dino_ablation} and \Cref{tab:dino_ablation_quan_gso_full}, respectively. All methods were evaluated on the IM~\cite{xu2024instantmesh} reconstruction pipeline.

\begin{table}[h]
    \centering
    \resizebox{0.6\columnwidth}{!}{
    \begin{tabular}{cccc}
     \toprule
     Method               & CD$\downarrow$  & F1$\uparrow$ & HD$\downarrow$   \\ \midrule
     \addlinespace[4pt] 
     $\text{SV3D}_{\text{IM}}$~\cite{voleti2024sv3d}    & 4.60 & 7.74& 18.0\\
     $\text{Ours}_{\text{CLIP}_{\text{IM}}}$ & \textbf{3.93} & \textbf{9.99} & \textbf{16.1}\\ 
     $\text{Ours}_{\text{DINO}_{\text{IM}}}$ & 3.97 & 9.91& 16.3\\
     $\text{Ours}_{\text{IM}}$ & \textbf{3.93} & 9.93& \textbf{16.1}\\
     \bottomrule
    \end{tabular}}
    \caption{Quantitative results of single-view 3D reconstruction on the full GSO dataset with static camera trajectory and adaptive trajectories with CLIP~\cite{radford2021learning}, DINOv2~\cite{oquab2023dinov2} and VGG16~\cite{simonyan2014very} as feature extractors, respectively. Numbers in bold are ranked first.}
    \label{tab:dino_ablation_quan_gso_full}
\end{table}

The results demonstrate that all feature extractors contribute to developing more effective camera paths, yielding improvements in both qualitative and quantitative metrics compared to static trajectories. Among the different feature encoder choices, VGG16 and CLIP exhibit similar performance with minimal differences in F1 scores, while DINOv2 falls slightly short.

\section{Coverage Metric}
We further evaluate static, random, and adaptive trajectories using a coverage metric. Given the ground truth (GT) mesh as input, we render the GT mesh at 21 camera poses for each trajectory type. The visible area in each render is marked as visible on the corresponding mesh texture UV map, and the final coverage metric is measured as the accumulated visible region over the entire texturable UV map area. Averaged across all objects in the GSO dataset, the static trajectory achieved an 87.8\% coverage rate, random trajectory 89.0\%, adaptive trajectory with CLIP 90.3\%, VGG16 90.5\%, and DINOv2 93.4\%.

While the coverage metric indicates the occlusion-revelation capability of different trajectories to some extent, we emphasize that it involves a trade-off with reconstruction quality. For example, using a greedy algorithm for extreme occlusion-revelation path-finding can achieve near 100\% coverage, but it induces drastic elevation changes and degrades video generation quality due to sudden viewpoint transitions. Considering both metrics, feature encoders including VGG16 and CLIP strike an effective balance between coverage and quality.

\section{Orbital Camera Trajectories}

As described in the main paper, we implement a closed-loop orbital camera trajectory to optimize viewpoint sampling. The camera orbit is structured into four quarters, with camera elevation changing at a constant rate within each quarter. Our path planning algorithm determines the increment steps for the first two segments, while the final two segments mirror these changes with negated values to complete the loop.

This design approach is motivated by two key considerations:
\begin{itemize}
    \item \textbf{Search Space Reduction:} Camera pose planning exists in a continuous space with virtually infinite candidate positions. By discretizing elevation changes, we significantly reduce the search space, enabling efficient execution of combinatorial optimization to determine an effective path.
    
    \item \textbf{Constraint Balance:} The closed-loop structure enforces zero total variation in elevation, creating a balance between exploring diverse elevation angles while remaining within the operational capabilities of the video diffusion model.
\end{itemize}

Our empirical observations indicate that unconstrained greedy algorithms---those without total variation constraints in elevation and closed-loop confinement---frequently produce steeply descending trajectories that result in distorted video generation outputs.

While we acknowledge the current limitations of our approach in exploring asymmetric objects, our method demonstrates significant improvements over static camera trajectories. We believe this work provides valuable insights for future research in reconstruction-centric camera path planning.

\section{Complementary Role in Direct 3D supervision}

Recent methods employing direct 3D supervision~\cite{xiang2024structured, zhang2024clay} have demonstrated superior performance in 3D generation and reconstruction, achieving high-quality results with faster inference than multiview-based approaches. To investigate this further, we conduct an ablation study on Trellis~\cite{xiang2024structured} comparing single image input against multiview inputs sampled from orbital camera trajectories. For multiview evaluation, we sample 6 views on static camera trajectories and 6 views on adaptive camera trajectories, forming three comparison groups: Trellis, Multi-Trellis Static, and Multi-Trellis Adaptive.

\begin{figure*}
    \centering
    \includegraphics[width=\linewidth]{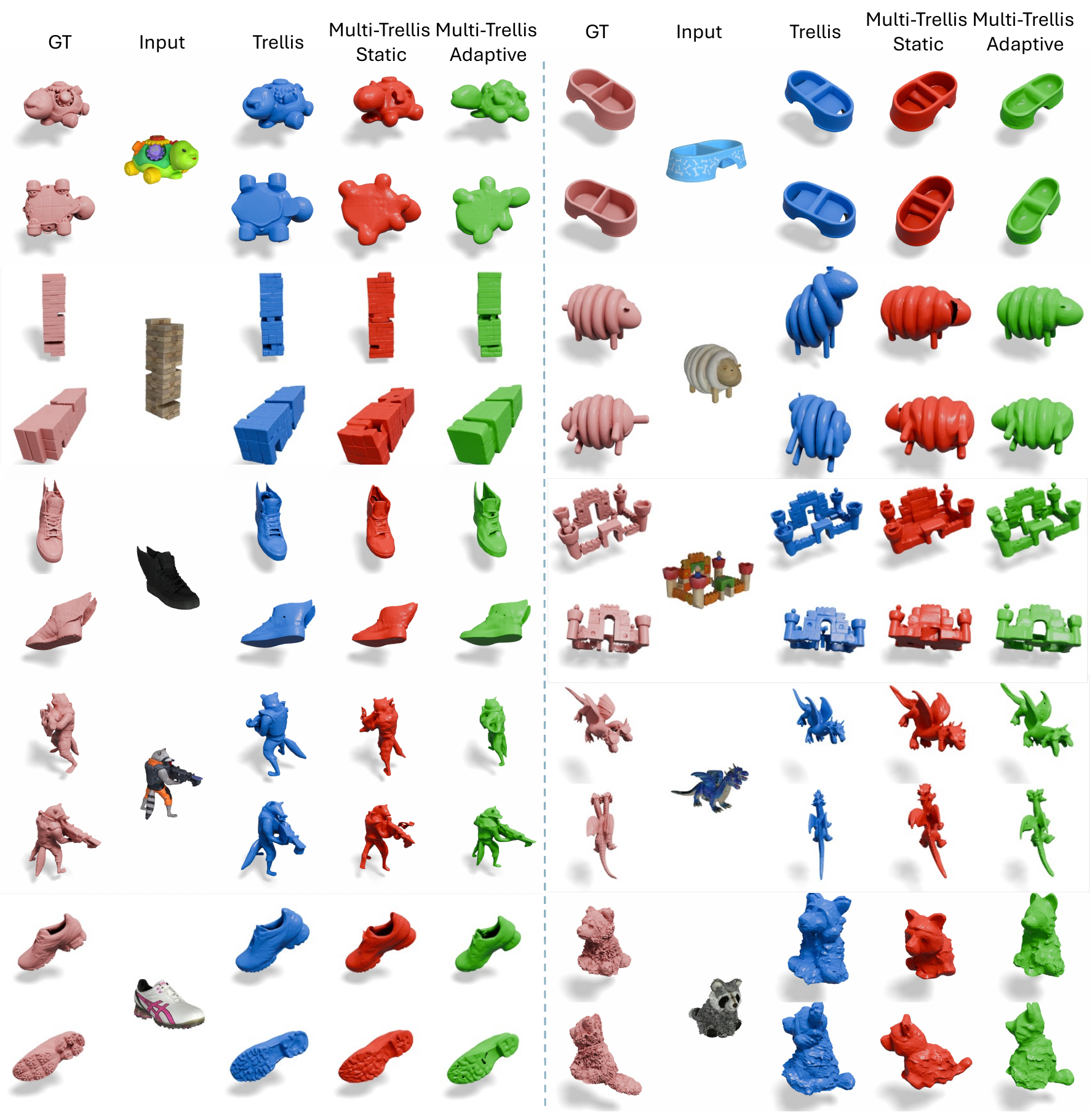}
    \caption{3D reconstruction results obtained using Trellis~\cite{xiang2024structured} with single image input and multiview image inputs sourced from static and adaptive camera trajectories. Single-view inputs consistently produce cleaner results with richer geometric details, as multiview inputs introduce additional blur from hallucinated views, as universally observed across all samples. However, multiview inputs demonstrate advantages for handling self-occlusion scenarios. Notable examples include R1 (\textbf{R}ight column, row \textbf{1}), R2, and R4, where the dog bowl is reconstructed with a solid base, the distorted sheep model is regularized, and the dual-head feature of the dragon is properly emphasized. These cases highlight how strategic multiview selection can enhance reconstruction quality in geometrically challenging scenarios despite the general trade-off in detail fidelity.}
    \label{fig:trellis-comparison}
\end{figure*}

The qualitative results in~\Cref{fig:trellis-comparison} reveal that multiview input Trellis does not consistently outperform its single-view counterpart. This degradation can be attributed to blur and distortions introduced by hallucinated multiview images, which compromise geometric detail quality as evidenced in side-by-side comparisons. However, when input views exhibit significant self-occlusion—as observed in rows 1, 2, and 4 of the right column in~\Cref{fig:trellis-comparison}—additional multiviews enhance geometric awareness in occluded regions, supporting more regularized and occlusion-aware reconstruction. Notably, adaptive multiviews consistently outperform static multiviews in these scenarios due to their superior occlusion-revelation capabilities.

While these advances in direct 3D supervision might suggest that multiview reconstruction is becoming obsolete, we contend that both paradigms offer complementary strengths. Although multiview inputs do not universally improve results, particularly when hallucinated views suffer from distortion or inconsistency, our findings indicate that multiview generation and direct 3D reconstruction can synergistically complement each other when views are strategically selected. We hope this exploration of adaptive trajectory planning will inspire further research into harnessing the combined potential of these two paradigms.



\end{document}